\documentclass[conference]{IEEEtran}
\IEEEoverridecommandlockouts
\usepackage{cite}
\usepackage{amsmath,amssymb,amsfonts}
\usepackage{graphicx}
\usepackage{textcomp}
\usepackage{xcolor}
\usepackage{xspace}
\usepackage{hyperref}
\usepackage{tabu}
\usepackage{algorithm}
\usepackage[noend]{algpseudocode}
\usepackage{array}
\usepackage[T1]{fontenc}
\usepackage{enumitem}

\newcommand{\WF}{{\textsc{Stitcher}}\xspace} 
\newcommand{\ST}{\WF} 
\newcolumntype{P}[1]{>{\centering\arraybackslash}p{#1}}
\newcolumntype{M}[1]{>{\centering\arraybackslash}m{#1}}

\def\BibTeX{{\rm B\kern-.05em{\sc i\kern-.025em b}\kern-.08em
    T\kern-.1667em\lower.7ex\hbox{E}\kern-.125emX}}

\usepackage{tikz,xcolor,hyperref}

\definecolor{lime}{HTML}{A6CE39}
\DeclareRobustCommand{\orcidicon}{%
	\begin{tikzpicture}
	\draw[lime, fill=lime] (0,0) 
	circle [radius=0.16] 
	node[white] {{\fontfamily{qag}\selectfont \tiny ID}};
	\draw[white, fill=white] (-0.0625,0.095) 
	circle [radius=0.007];
	\end{tikzpicture}
	\hspace{-2mm}
}

\foreach \x in {A, ..., Z}{%
	\expandafter\xdef\csname orcid\x\endcsname{\noexpand\href{https://orcid.org/\csname orcidauthor\x\endcsname}{\noexpand\orcidicon}}
}

\begin{document}

\title{\ST: Correlating Digital Forensic Evidence on Internet-of-Things Devices\\
}

\author{\IEEEauthorblockN{Yee Ching Tok\orcidA{}}
\IEEEauthorblockA{\textit{Singapore Univ. of Tech. and Design} \\
Singapore \\
yeeching\_tok@mymail.sutd.edu.sg}
\and
\IEEEauthorblockN{Chundong Wang\IEEEauthorrefmark{1}\orcidB{}}
\IEEEauthorblockA{\textit{ShanghaiTech University}\\
\thanks{*This work was done when Chundong worked in Singapore University of Technology and Design.}
China \\
wangchd@shanghaitech.edu.cn}
\and
\IEEEauthorblockN{Sudipta Chattopadhyay\orcidC{}}
\IEEEauthorblockA{\textit{Singapore Univ. of Tech. and Design} \\
Singapore \\
sudipta\_chattopadhyay@sutd.edu.sg}
}

\maketitle
\thispagestyle{plain}
\pagestyle{plain}

\begin{abstract}
The increasing adoption of Internet-of-Things (IoT) devices present new challenges to digital forensic investigators and law enforcement agencies when investigation into cybercrime on these new platforms are required. However, there has been no formal study to document actual challenges faced by investigators and whether existing tools help them in their work. Prior issues such as the correlation and consistency problem in digital forensic evidence have also become a pressing concern in light of numerous evidence sources from IoT devices.

Motivated by these observations, we conduct a user study with 39 digital forensic investigators from both public and private sectors to document the challenges they faced in traditional and IoT digital forensics. We also created a tool, \ST, that addresses the technical challenges faced by investigators when handling IoT digital forensics investigation.

We simulated an IoT crime that mimics sophisticated cybercriminals and invited our user study participants to utilize \ST to investigate the crime. The efficacy of \ST is confirmed by our study results where 96.2\% of users indicated that \ST assisted them in handling the crime, and 61.5\% of users who used \ST with its full features solved the crime completely.
\end{abstract}


\section{Introduction}\label{Introduction}

Security is of paramount importance for all digital systems and the emergence of Internet-of-Things (IoT) technology has attracted new adopters and adversaries. It is reported that attacks on IoT devices have increased by more than 300\% in the first half of 2019~\cite{WorldEconomicForum_2020}, demonstrating that IoT has become a favorite target by advanced adversaries and cybercriminals alike. The financial damage caused by cybercrime has been estimated to hit US\$6 trillion by the year 2021~\cite{WorldEconomicForum_2020}. Thus, it is imperative that Digital Forensics Investigators (DFI) and/or Law Enforcement Agencies (LEA) can address and investigate these emerging forms of cybercrime, just as how they handle such forms of cybercrime on traditional computer systems by performing digital forensics and analyzing digital evidence. However, the domain of IoT digital forensics is immature and requires further work. For example, there has been research on identifying some sources of evidence that can be retrieved from IoT devices~\cite{SERVIDA2019S22}. Unfortunately, inadequately trained DFI may not be familiar with how IoT devices and smart homes are structured, hence affecting how IoT evidence types could be classified and presented in reports or courts of law. This could cause unnecessary confusion to judges, lawyers or clients as there is no consistent way in which evidence are presented. The heterogeneity of IoT devices also brings in an increased amount of evidence sources for DFI and/or LEA to correlate. 

It is not trivial to determine the root cause of issues mentioned in the preceding paragraph. It could be due to a lack of awareness by device manufacturers who did not design their products securely, hence resulting in security issues such as leaving private keys embedded in firmware~\cite{CVE-2017-13663}. Moreover, a lack of user awareness and device support to patch vulnerabilities in IoT devices provide cybercriminals ample opportunities to compromise such devices. Technical challenges faced by DFI or LEA when they have to investigate IoT related crimes play a major role as well. The inability to effectively investigate IoT related crimes and attacks only encourages adversaries and cybercriminals to exploit such opportunities. Gaining financial rewards while committing a crime, and yet not be prosecuted or identified in a timely fashion is attractive for adversaries and organized crime syndicates.

While research in digital forensics on IoT devices is growing, there are still gaps in capabilities required by DFI. Prior research has identified challenges in the digital forensics domain~\cite{Raghavan2013Dfrc}, but there has been no formal survey directed towards DFI that investigate the challenges they face, especially in the area of IoT digital forensics. Although some existing work addresses the problem of automating forensic evidence discovery and correlation~\cite{CASE2008S65}, it was limited only to traditional computer systems rather than IoT devices. 

Based on the issues we have highlighted, we conducted a survey and user study with 39 DFI from both public and private sectors to concretely and comparatively identify current challenges they face in both traditional and IoT digital forensics. The identified challenges will serve as a guideline for possible future research work on IoT digital forensics. Additionally, stakeholders who are looking into creating an IoT digital forensics capability in their respective organizations can utilize the identified challenges to plan or obtain the necessary resources to secure success.  Finally, the challenges also motivated us to create a tool, \ST. \ST is an automated evidence classification and correlation tool designed to assist DFI in IoT digital forensics. When retrieved IoT evidence such as firmware images, network packet captures and system processes are provided to \ST, it classifies the evidence based on a combination of relevant international ISO standards (ISO 27050-1:2019~\cite{information_technology_security_techniques_electronic_discovery_2019} and ISO 30141:2018~\cite{IoT_reference_architecture_2018}). \ST then processes the evidence and correlates matching data points within the evidence, and finally outputs the results to DFI. With the assistance of \ST, DFI are empowered to classify and analyze multiple evidence sources in a shorter amount of time, and thus solving cases in a more timely fashion.

The contributions of our research are summarized as follows:
\begin{enumerate}
    \item We present the current forensic challenges faced by 39 DFI coming from both public and private sectors. The survey results consists of 1) current challenges faced in traditional digital forensics, 2) challenges faced in IoT digital forensics and 3) challenges faced in a simulated IoT crime that mimics sophisticated cybercriminals.
	\item We develop and present a tool, \ST, aimed to assist DFI in mitigating challenges they face in IoT digital forensics. \ST will aid DFI in classifying, processing and correlating IoT forensic evidence. Baseline data, or known good data are also accepted to strengthen correlation between evidence sources; and
	\item We demonstrate the efficacy of \ST by presenting the results of our user study participated by 39 DFI. In the user study, DFI were invited to use \ST to classify, process and correlate digital evidence retrieved from a simulated IoT crime that mimics sophisticated cybercriminals. The efficacy of \ST is demonstrated by our study results where 96.2\% of users indicated that \ST assisted them in handling the crime, and 61.5\% of users who used \ST with its full features solved the crime completely.

\end{enumerate}

For reproducibility and advancing the research in IoT digital forensics, our tool and experimental data are publicly available at: \url{https://github.com/poppopretn/Stitcher}.

The rest of this paper is organized as follows. In Section~\ref{Background}, we present the background of the paper. In Section~\ref{InfraStudyDesign}, we detail the design of our experimental infrastructure and user study. In Section~\ref{Tool}, we show the design of \ST. In Section~\ref{StudyResults}, we present the results of our user study. In Section~\ref{Limitations}, we highlight the limitations of our research and \ST. In Section~\ref{RelatedWork}, we summarize current related work. Finally, we conclude the paper in Section~\ref{Conclusion}. 
\section{Background} \label{Background}

IoT devices are getting increasingly popular and ubiquitous in homes, offices and critical infrastructure such as water and electricity plants. The adoption of IoT devices has led to increased productivity and potential cost savings, but also presents a new paradigm of challenges to DFI and LEA as such devices become a new target for criminal activities. Cybercriminals and state-sponsored actors could attempt to backdoor devices via their update mechanisms~\cite{Jacob:2017:CFS} or exploit design flaws in devices to perpetrate attacks~\cite{CVE-2017-13663, CVE-2017-13664}. 

The digital forensics landscape appears to be getting by with the plethora of open-source and commercial tools available to DFI and LEA to use such as \textit{Volatility}~\cite{Volatility}, \textit{Rekall}~\cite{Rekall}, \textit{Autopsy}~\cite{Autopsy}, \textit{The Sleuth Kit}~\cite{TheSleuthKit}, \textit{AccessData Forensic Toolkit (FTK)}~\cite{AccessData} and \textit{Encase Forensic}~\cite{EnCase}. Moreover, there are also valuable guidance of digital forensic and various artifacts on traditional computer systems~\cite{Casey2011}. However, these resources were mostly intended for traditional digital forensics. There are multiple difficulties that have plagued DFI in their investigation efforts, especially in emerging technologies. From the user study we conducted (outcomes are discussed in detail under Section~\ref{RQ3}), major challenges highlighted by DFI include a lack of training/knowledge in emerging technologies, multiple evidence sources to examine and correlating those evidence sources.

\subsection{Types of IoT Evidence} \label{TypeOfIoTEvidence}

The success of IoT digital forensics is determined by the availability of accurate evidence sources that can be examined. Evidence examination consists of preparing storage mechanisms to contain extracted evidence, extraction of relevant evidence (physical or logical), analyzing evidence for data points of interest and finally obtaining a conclusion~\cite{NIJ_2004}. DFI should be mindful of factors that can affect the accuracy of evidence generated by IoT devices, such as time zone settings and storage location of various log files. To address these factors, accuracy can be improved by cross-referencing different evidence sources, and by ensuring that relevant evidence sources needed for investigation are identified and collected. For instance, network activity of an event of interest recorded in network packet captures can be utilized to correlate log files that reflect the same event, thus confirming accuracy of the log files~\cite{Casey2011}. As part of our research contribution, we identified possible evidence sources (see Sections~\ref{FirmwareImage} to~\ref{SystemProcesses}) that could be examined by DFI for investigation.

\begin{figure}[th]
\centering
\includegraphics[width=0.8\columnwidth]{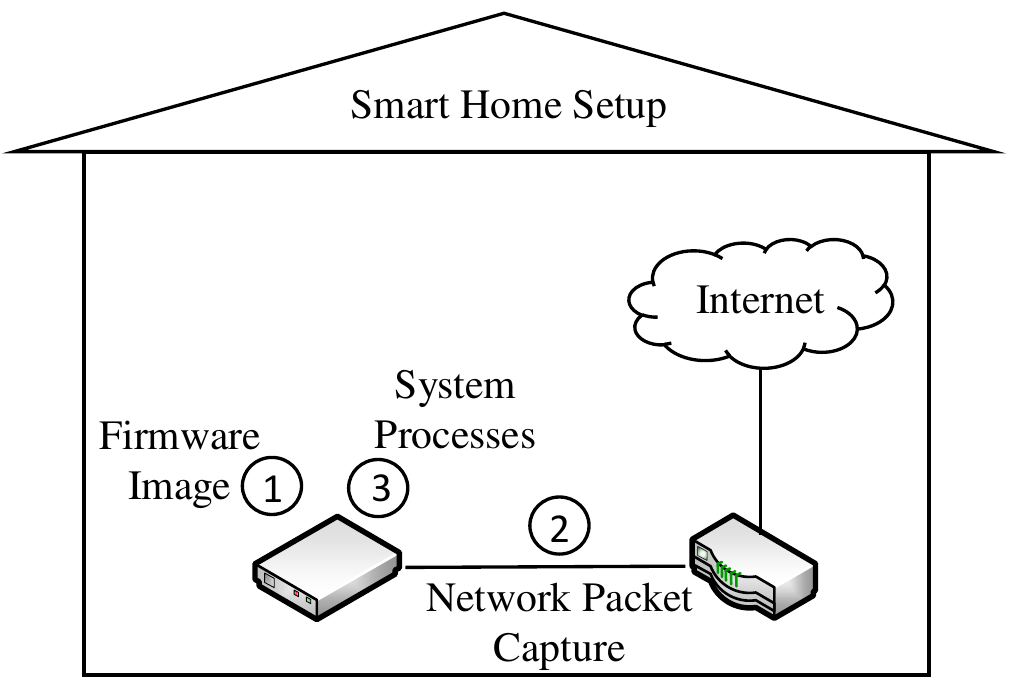}
\caption{Common IoT Setup Scenario in Smart Homes}
\label{fig:IoT_Setup}
\end{figure}

While evidence stored on cloud services could also be used, we did not want to focus on them as access to these evidence could be challenging and time-consuming due to legal paperwork. As such, we only focused on evidence that could be retrieved from the crime scene or system owners. As illustrated in~\autoref{fig:IoT_Setup}, we identified three sources of evidence that are useful during IoT digital forensics - \raisebox{.5pt}{\textcircled{\raisebox{-.9pt} {1}}} firmware image, \raisebox{.5pt}{\textcircled{\raisebox{-.9pt} {2}}} network packet capture and \raisebox{.5pt}{\textcircled{\raisebox{-.9pt} {3}}} system processes.

\subsubsection{Firmware Image} \label{FirmwareImage}
Firmware images are purpose-built software operating systems that give embedded devices their functionality. However, unlike conventional operating systems, firmware images are usually compressed and have to be extracted by specialized software tools such as \textit{binwalk}~\cite{binwalk}. Depending on whether it is extracted directly from the device or provided as-is by the manufacturer, firmware images may contain data that could provide additional context, such as logs or verbose debug files. When firmware images are unpacked and extracted, they yield multiple files and directories just like a traditional operating system such as Ubuntu Linux or Microsoft Windows. As such, the approach to examine firmware artifacts is similar to that of a forensic disk image obtained in a traditional digital forensic case. DFI would usually be interested in artifacts such as file names, list of file directories, contents of the file itself (text strings within files) and file hashes (unique alphanumeric characters generated by hashing algorithms). However, the process to unpack firmware images could be confusing and technically complex. Moreover, there could be some loss of context if the firmware image was examined statically. For example, if the firmware was obfuscated, DFI may find it challenging to uncover its original functionality just by static analysis.

The firmware image of an embedded device can be extracted via serial connection, Joint Task Action Group (JTAG) interface or downloaded from manufacturers' support sites (if they are hosted there). Last but not the least, firmware images can be retrieved from network packet captures if a firmware upgrade was performed and the network was configured to capture network traffic.

\subsubsection{Network Packet Capture}
Network traffic is usually not captured due to storage limitations and privacy concerns. Moreover, modern network traffic is often encrypted end-to-end, and thus may require user authentication and more implementations such as Secure Sockets Layer (SSL) decryption. Despite the minor inconveniences, network packet captures offer a wealth of information that can be utilized to accelerate an investigation as the majority of communications happen over the network. Useful information such as source and destination IP addresses, hardware addresses, source and destination port numbers, protocols and data payloads are traceable and retrievable from network packet captures. By having visibility of what had happened in the network, important artifacts such as Command and Control (C2) traffic from backdoored devices or malicious traffic targeted towards devices can be observed. 

For more advanced attack vectors such as self-deleting malware after establishing a connection or exfiltrating data, their presence can still be inferred from captured network traffic despite the absence of the original file. Investigating port numbers could yield useful information as using the repeated use of a same port number might indicate the presence of a C2 channel. Additionally, well known default port numbers such as port 4444 used by exploitation frameworks like \textit{Metasploit}~\cite{metasploit} could be revealed as well.

Network packet captures can be obtained by enabling port mirroring on networking devices and plugging in hardware sniffers such as network taps. Alternatively, commercial solutions such as Gigamon~\cite{Gigamon} and Riverbed~\cite{Riverbed} can provide these types of network capture capabilities as well.

\subsubsection{System Processes} \label{SystemProcesses}
To execute their functionality, IoT devices need to execute programs embedded within the firmware, and thus start the corresponding processes that run those programs. By examining the list of processes that were active at the point of time of extraction, it is possible to obtain a snapshot of what has been exactly happening on the IoT device. For example, by comparing the list of processes with a clean build provided by the manufacturer, it is possible to ascertain foreign processes that are running.

The retrieval of system processes will require DFI to directly interact with the device and run bash commands such as \textit{ps}. While there could be some misgivings about directly interacting with a system under investigation, it should be noted that memory forensics on traditional computer systems also required interaction with the system to obtain more information. As long as proper documentation of actions and the resulting evidence generated were kept, DFI will be able to have insights on what could have transpired on the IoT device.

\subsection{Correlation and Consistency Problem}
DFI face the challenge of correlating multiple sources of evidence, and also analysing the evidence for consistency~\cite{Raghavan2013Dfrc}. With the growth in usage of IoT devices, DFI can anticipate efforts required for correlation and consistency of digital evidence to grow exponentially when cybercrime occurs on IoT devices.

This presents a huge administrative and technological gap for DFI and LEA when they are faced with cybercrimes involving IoT devices. For example, traditional digital forensic techniques such as disk imaging and live memory analysis are applicable for investigation on cybercrime for conventional computer systems. Such techniques are also possible on smartphones via specialized hardware and software tools such as \textit{Cellebrite}~\cite{Cellebrite}. However, most IoT devices have components soldered together without standard ports present on computers and smartphones. In most cases, access to serial or JTAG interfaces often require disassembly of the IoT device. As such, traditional digital forensic techniques highlighted earlier cannot be executed. The challenges are further exacerbated by unfamiliarity with IoT devices due to the heterogeneity of devices and potentially large amounts of evidence sources to be examined, leading to longer investigation efforts, gaps in evidence collection and spiralling backlogs of cases to investigate. Standard evidence documentation, procedures and investigation workflows that depend on such techniques will thus also become invalid.

\subsection{Addressing Challenges in IoT Digital Forensics}

DFI in LEA or private companies trawl through literature to properly classify multiple evidence sources in an internationally agreed convention. As IoT is an emerging area, clients and legal courts may struggle to understand the architecture of an IoT set-up. Without a proper nomenclature, valuable evidence may be at risk of being inadmissible in a court of law. 

Based on the scenario highlighted above, suppose a tool has been developed to address the challenges. The DFI is provided with a number of IoT evidence sources to examine and executes the said tool. The provided evidence is classified based on internationally agreed standards which will aid him/her in reducing time required for report writing later and reducing ambiguity if presentation of evidence in a court of law is required. After evidence classification, the evidence is parsed so that the correlation algorithms can correlate various data points that appear consistently in the evidence sources. Finally, the correlation algorithms correlate the various data points that are related and present them to the DFI. After examining the output, the correlated data points in various evidence sources help the DFI stitch together seemingly disparate data points and reveal the crime that occurred, leading to the resolution of the case and provision of a satisfactory report promptly.

Suppose we now apply the same scenario based on current state of the art tools and methodologies. DFI have to manually examine the various IoT evidence sources that were provided. A long time will be required if a particular evidence source contains a large number of files and the DFI will have to rely on personal experience if no further details about the crime are provided. All evidence sources have to be cross examined for data points that turn up and cross referenced to see if any potential criminal activity occurred, which could take a few days to complete. Finally, as IoT is an emerging area and there has been no prior cases, the DFI will require much more time (as compared to the situation in the previous paragraph) to solve the case and provide a satisfactory report.
\section{Infrastructure and Study Design} \label{InfraStudyDesign}

We present the design of our infrastructure and study that would be instrumental in our experiments and research.
\subsection{Research Questions (RQ)} \label{ResearchQn}

We had two main objectives in the User Study. Firstly, we wanted to profile the DFI work experience and training received, along with challenges they faced in traditional and IoT digital forensics. Secondly, we wanted to evaluate the performance of a software tool - \ST\  - developed to address new and existing challenges faced by DFI in the domain of IoT digital forensics. These objectives were translated into the following research questions:

\paragraph*{\textbf{RQ1 Background Knowledge}}How much does prior knowledge (such as material learned in Institutions of Higher Learning (IHL), professional training, and work experience) help DFI in forensics investigations, especially if they are faced with new frontiers of digital forensics, such as IoT digital forensics?

\paragraph*{\textbf{RQ2 Challenges in Traditional Digital Forensics}}  What are the top challenges and concerns faced by DFI in the field of traditional digital forensics right now? 

\paragraph*{\textbf{RQ3 Challenges in IoT Digital Forensics}} What are the top gaps, challenges and concerns DFI face in emerging areas, such as digital forensics on IoT devices?

\paragraph*{\textbf{RQ4 DFI Performance in IoT Digital Forensics}} How do DFI perform when handling new cases such as IoT digital forensics?

\paragraph*{\textbf{RQ5 Empowering DFI}} Would the introduction of a software tool designed to address the gaps make DFI more efficient and effective in handling IoT digital forensics? 

\subsection{Objects and Infrastructure}
In this section, we present the objects and infrastructure that are utilized in the User Study.

\subsubsection{Objects}

The objects under study consists of multiple objects that are broadly classified into the following two categories:
\begin{enumerate}[label=\roman*)]
	\item \textbf{Tool:} We developed a software tool - \ST. To evaluate the effectiveness and efficiency of the tool, we deliberately built three configurations of the tool. The first configuration was designed for the control group and to evaluate the current digital forensics landscape. The second configuration had reduced functionality, while the third configuration of the tool had all features enabled. A more detailed explanation of \ST can be found at Section~\ref{Tool}.
	\item \textbf{Digital Evidence:} We designed and implemented an attack scenario that replicates a plausible way malicious actors could use to compromise a conventional IoT setup. In this study, the attack scenario chosen was the introduction of a backdoor into the firmware image, thus allowing malicious actors to communicate and issue commands to the IoT device. Moreover, the backdoor is able to survive reboots and start itself again. We took on the role of malicious actors, connected to the device using netcat and simulated interaction with the device by typing in some Linux commands. The following three types of digital evidence were obtained from the setup before and after the compromise: firmware image, network packet capture, and process list of the IoT device (iSmartAlarm CubeOne).
\end{enumerate}

\subsubsection{Infrastructure}

The infrastructure under study consists of specially designed infrastructure as follows:

\begin{figure}[b]
\centering
\includegraphics[width=0.8\columnwidth]{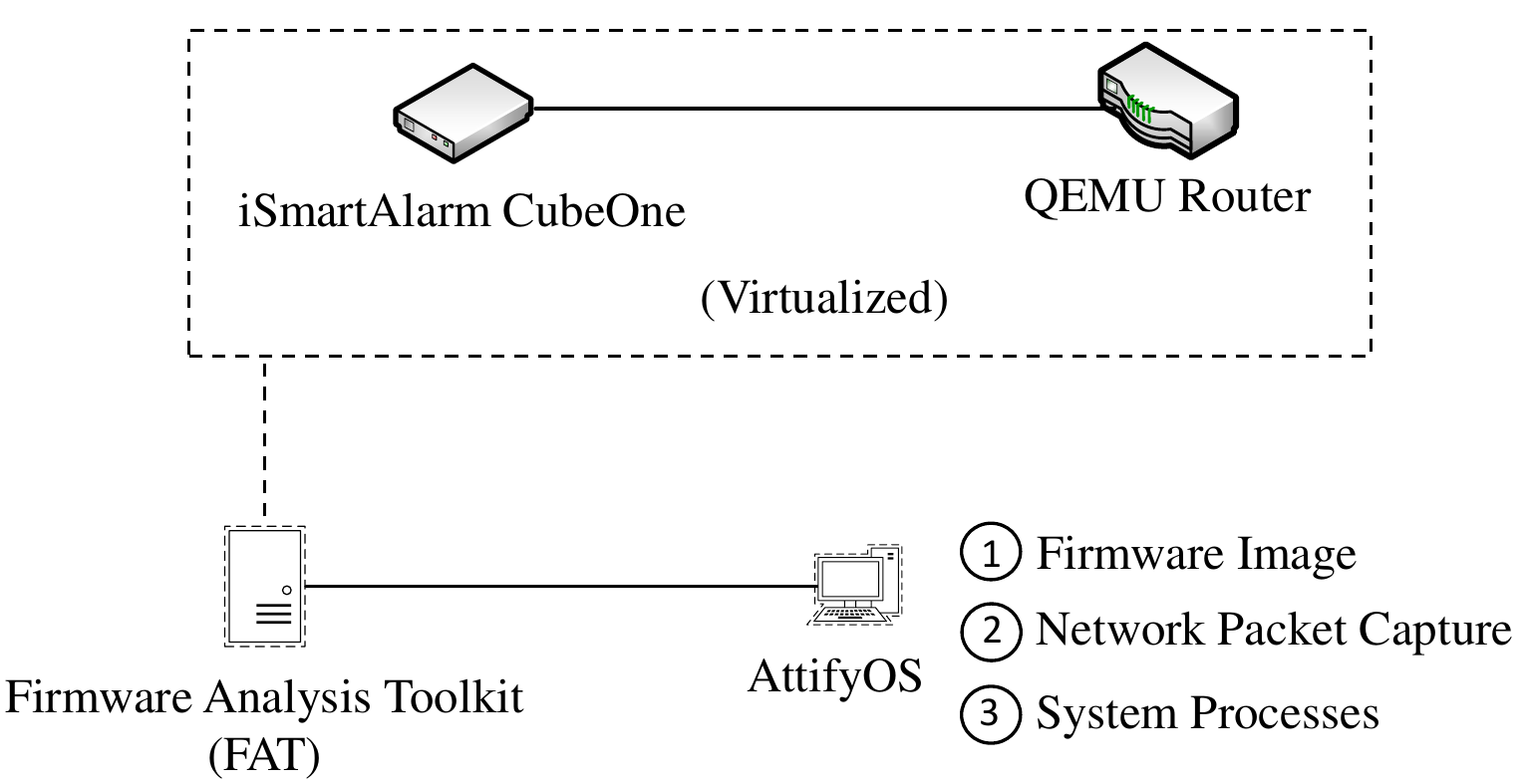}
\caption{Infrastructure of Scenario}
\label{fig:SI}
\end{figure}

\begin{enumerate}[label=\roman*)]
	\item \textbf{Scenario Infrastructure:} With reference to~\autoref{fig:SI}, we utilized a Linux Ubuntu distribution, AttifyOS~\cite{AttifyOS}, to virtualize the firmware of iSmartAlarm CubeOne (version 2.2.4.11) via Firmware Analysis Toolkit (FAT)~\cite{FAT}. Using a virtualized infrastructure offers multiple benefits. One advantage is that it reduces the complexity of firmware modifications as an image file was created by FAT when the firmware was virtualized. Virtualization also allowed us to avoid potentially bricking the physical device in our pursuit to create the attack scenario. This infrastructure was only used to generate the evidence (firmware image, network packet captures and process list) required for the pilot study (see Section~\ref{PilotStudy}) and user study (see Section~\ref{UserStudy}).
	
	\textbf{Backdooring iSmartAlarm CubeOne.} By mounting and modifying the image file used by FAT when virtualizing the firmware, we tampered with the firmware image directly. An existing backdoor program written in C language~\cite{CBackdoor} was modified and implanted into the firmware. We also modified the \textit{rcS} file located inside the \textit{/etc\_ro/} directory to ensure backdoor persistence is established upon booting. 
	
	\textbf{Evidence Retrieval.} To obtain the firmware image, we copied the firmware image file used by FAT. The network packet capture was obtained by running Wireshark on the network interface of AttifyOS. The process list was retrieved by connecting to the virtualized firmware from the backdoor connection and executing the \textit{ps} command of Linux to list all running processes. The output was then saved in a text file output.
	
	\item \textbf{Investigation and Analysis Infrastructure:} The investigation and analysis infrastructure consists of a Virtual Machine (VM) running Ubuntu 18.04.1 as a base operating system. Three VMs were created - 
	each having the evidence of the attack scenario generated from the scenario infrastructure and running a different configuration of \ST for the purposes of the pilot and user study. 
	The VMs were loaded individually on different host machines to be used for the pilot study and user study.
\end{enumerate}

\subsection{Pilot Study} \label{PilotStudy}

Pilot studies are vital in identifying blind spots and verifying that the main study is directed to the right audience. It also ensures that tasks designed for the main study are clear to participants. We ensured that the pilot study candidate had relevant industry experience and prior knowledge before commencing the pilot study.

During the pilot study, we first conducted a 20 minutes oral briefing to the candidate to explain the goals of the user study. Following that, the candidate was given 30 minutes for each configuration of the tool and attempted to solve the attack scenario. The candidate was observed closely while solving the scenario with the different configurations of the tool. Next, the candidate was presented with a survey to answer targeted questions with respect to Section~\ref{ResearchQn} and provide feedback on the tool. Through the pilot study, we are able to confirm that our scenario was realistic enough and refined our processes by designing an answer sheet for future participants to fill in their answers.

\subsection{Main Study: DFI} \label{UserStudy}

In this section, we detail the steps taken to conduct the user study.

\paragraph*{\textbf{Recruitment and Selection}} As skilled DFI are particularly rare, we relied on industry and personal connections to recruit suitable candidates. The candidates were a mix of DFI from both public and private sectors based in Singapore. We informed our point-of-contacts that interested participants should possess at least 1 of the the 3 following requirements:
\begin{enumerate}
    \item Studied digital forensics in an IHL.
	\item Took professional training and/or possess digital forensics certifications (e.g. GCFE, GCFA, GNFA, GREM, eCDFP, EnCE, ACE, etc).
	\item Investigated cases where digital forensics was required.
\end{enumerate}

A total of 39 DFI were recruited for the user study.

\paragraph*{\textbf{Tasks}}
We had a maximum of three participants per session as this allowed us to observe the DFI whilst they were trying to solve the scenario. In contrast to the pilot study, participants were given an extra 10 minutes writing time to fill up the answer sheet, though they were encouraged to document any notable observations during the first 30 minutes time frame. We also ensured that participants did not know there were actually three different configurations of \ST so as to obtain the best data possible.

\paragraph*{\textbf{Debriefing}}
The debrief consisted of how the scenario was constructed and what essential evidence should have been discovered from the evidence sources given. We highlighted the main differences of each configuration, and emphasized that the user study was not meant to formally evaluate technical competency. Finally, we asked the user study participants to fill in an online questionnaire indicating their background in digital forensics, challenges faced in their work and on IoT forensics, and provide feedback on the tool used (where applicable, as at least a third of the participants did not use the tool due to the nature of our study design).

\subsection{Measures of Investigation Performance by DFI}
We created the following categories that measured the performance of participating DFI: 

\begin{enumerate}
    \item \textbf{Full Solve.} Correctly identified most, if not all, of the evidence from the given evidence sources pertinent to the scenario. DFI were able to accurately highlight the particular attack scenario, write a brief report that captures the factors causing the given attack scenario, supported by the evidence sources that were identified.
	\item \textbf{Partial Solve.} Correctly identified at least half of the evidence from the given evidence sources pertinent to the scenario. DFI depicted a partially wrong attack scenario, but was able to write a brief report that depicts the factors that caused attack scenario, supported by the evidence sources that were identified. 
	\item \textbf{No Solve.} Correctly identified less than half of the evidence from the given evidence sources pertinent to the scenario. DFI depicted a wrong attack scenario, and/or unable to write a brief report that depicts the factors that caused attack scenario. DFI was also unable to process the raw evidence to establish the circumstances of the attack scenario. 
\end{enumerate}

DFI were categorized based on the answer sheet that was submitted and visual observations while they were participating in the user study. To further ensure accuracy of DFI categorization, answer sheets were reviewed at least twice after submission by a researcher who possessed all three requirements that user study participants must fulfil before being eligible to participate in the user study.

\subsection{Assumptions and Limitations}

We made a few assumptions and limitations for the study design, but these do not impact the usability of the tool or results of the study. They are as follows:
\begin{enumerate}
    \item The scenario environment was set up to facilitate the execution of forensic actions to retrieve evidence.
    \item We did not require DFI to ensure they had proper chain of custody of evidence as their primary role was to interact with the evidence that was provided.
    \item We assumed that little effort is required to retrieve the process list of the firmware that was running before and after the compromise. The IoT device must be able to support direct communication, either via serial or port connection. In our case, we made use of the established backdoor connection to obtain the process listing of the compromised firmware. As we were unable to directly connect to the iSmartAlarm CubeOne without tampering the firmware itself, we simulated the process list of a clean firmware by removing the backdoor process from the process list obtained when the firmware was backdoored. The accuracy of process list of a clean firmware is still maintained - we had investigated the device thoroughly before embarking on research and observed no visible backdoors on the original firmware. Since we tampered with the firmware and already had visibility on the additional system process we created, we are confident that the system process list recreated in this way best represents the processes that were running on an untampered firmware.
	\item We assumed that all evidence was retrieved in a forensically sound manner as \ST does not verify the integrity of evidence provided.
	\item We virtualized the IoT device as this was another method we discovered which could allow us to tamper with the firmware image in the absence of a zero-day vulnerability. 
	As such, while it is a slightly unconventional way to generate the corresponding IoT evidence (firmware image, network packet capture and system processes), we do not believe it has caused any bias in the tool evaluation as DFI work directly with the evidence sources in our user study. Even if we utilized a zero-day vulnerability (or a few zero-day vulnerabilities) to generate the same evidence sources, DFI are still going to use \ST to process the evidence sources in the same manner as described.
\end{enumerate}
\section{Tool} \label{Tool}

\begin{figure}[H]
\includegraphics[width=\columnwidth]{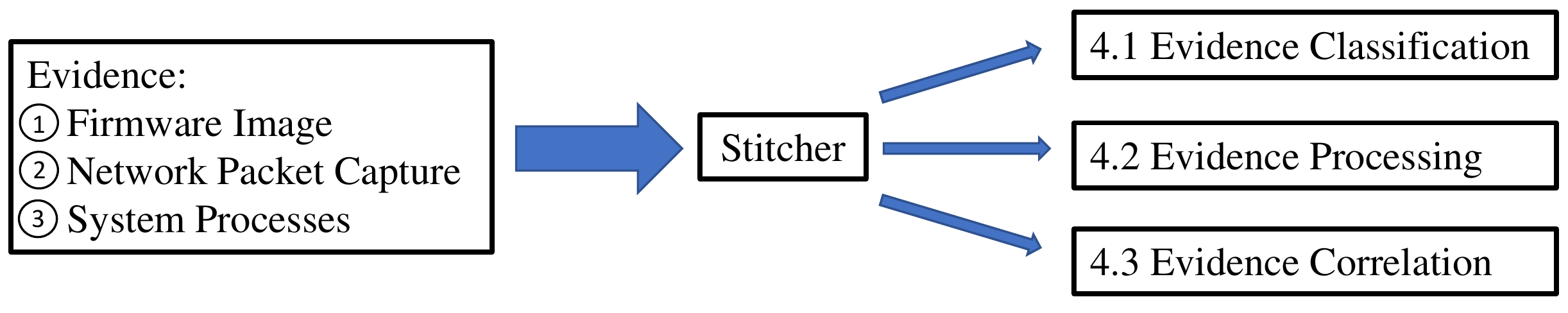}
\vspace{-18pt}\caption{Components of \ST}
\label{fig:ComponentsofST}
\end{figure}

With reference to~\autoref{fig:ComponentsofST}, \ST operates in three main ways - classifying evidence, processing evidence sources to facilitate correlation, and correlating processed evidence sources. 

\subsection{Evidence Classification}

As digital forensics on IoT systems is an emerging field, a standardized evidence classification scheme reduces ambiguity and provides a globally agreed structure for evidence presentation and discussion.

\begin{table} [H]
\centering
\caption{Evidence Classification using ISO27050:2019 and ISO30141:2018}
\begin{tabu} to 0.9\columnwidth { | X[l] | X[l] | X[l] | }
 \hline
 \vspace*{1pt}\textbf{Types of Evidence handled by \ST\vspace*{4pt}} & \vspace*{3pt}\textbf{Classification via ISO27050-1:2019} & \vspace*{3pt}\textbf{Classification via ISO30141:2018} \\
 \hline
 \vspace*{1pt}Firmware Image (Section 4.2.1) & \vspace*{1pt}7.2.2 Active data 
7.3.2 Custodian data source

7.4.2 Native format\vspace*{6pt}  & \vspace*{1pt}8.2.3.9 Data store  \\
 \hline
 \vspace*{1pt}Network Packet Capture (Section 4.2.2)  & \vspace*{1pt}7.2.3 Inactive data 
7.3.3 Non-custodian data source

7.4.2 Native format\vspace*{6pt} & \vspace*{1pt}8.2.3.8 Network  \\
 \hline
 \vspace*{1pt}System Processes (Section 4.2.3)  & \vspace*{1pt}7.2.2 Active data
7.3.2 Custodian data source

7.4.2 Native format\vspace*{6pt} & \vspace*{1pt}8.2.3.5 Service  \\
\hline
\end{tabu}
\label{table:1}
\end{table}

\begin{figure}[th]
\includegraphics[width=\columnwidth]{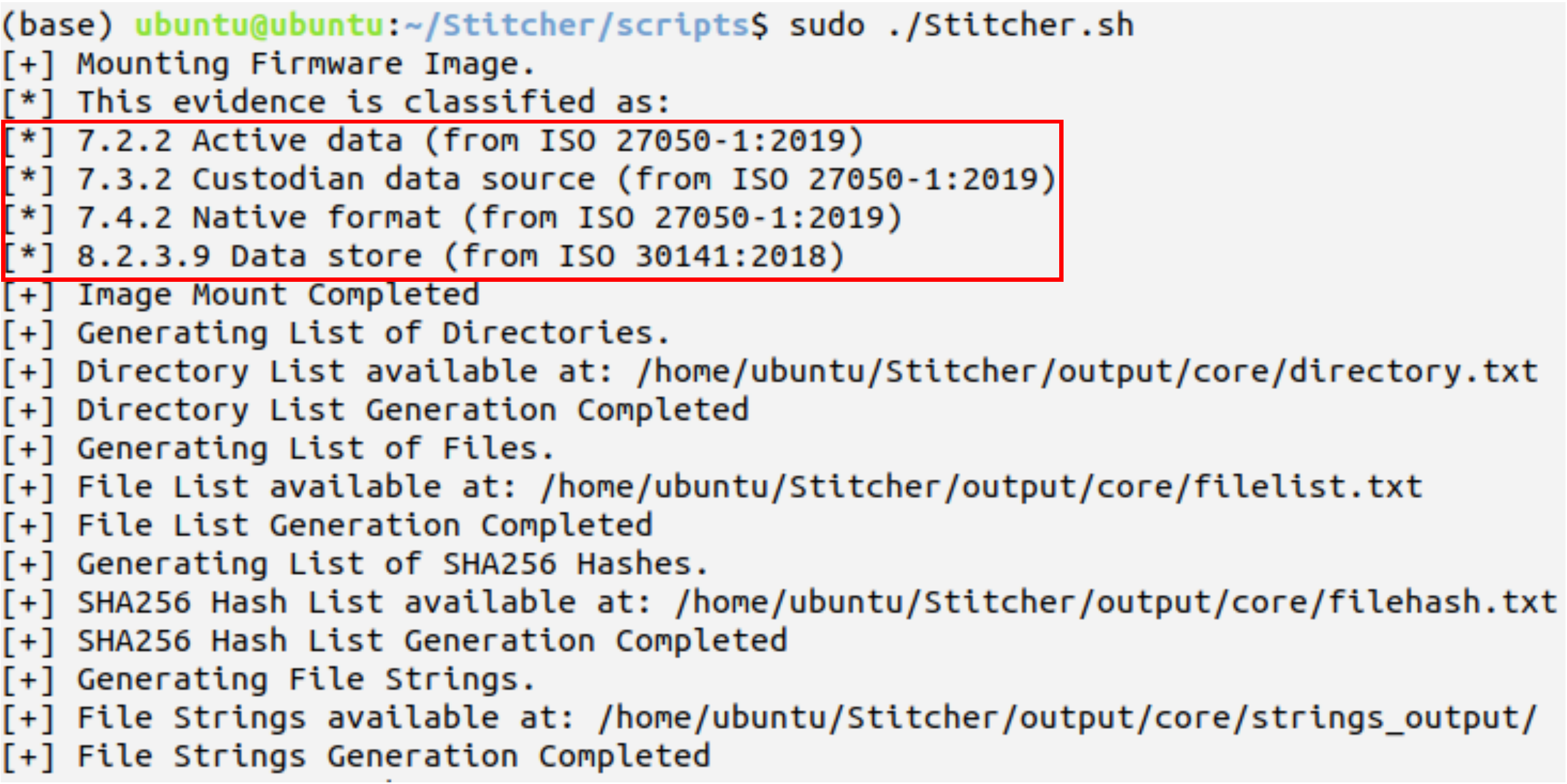}
\vspace{-18pt}\caption{Evidence Classification of \ST}
\label{fig:STClassifcation}\vspace{-10pt}
\end{figure}

With reference to~\autoref{table:1}, the evidence classification is based on two ISO standards - ISO27050-1:2019~\cite{information_technology_security_techniques_electronic_discovery_2019} and ISO30141:2018~\cite{IoT_reference_architecture_2018}. ISO27050-1:2019 has been widely used as a means to classify traditional digital evidence. However, ISO27050-1:2019 was primarily developed to classify evidence of conventional computer systems and is inadequate when IoT evidence is presented. For example, evidence gathered from components such as RAM or hard disks can be directly referenced to ISO27050-1:2019 (residual data and active data respectively). However, IoT evidence cannot be directly mapped to ISO27050-1:2019 without losing context due to heterogeneity of devices, deployment scenarios and possible types of evidence that can be retrieved. ISO30141:2018 addresses this gap as it offers multiple views of how an IoT device is located in the network and interacts with other components of an IoT setup. Coverage provided by these ISO standards helps with reporting despite the heterogeneity of IoT devices (as shown in~\autoref{fig:STClassifcation}).

\subsection{Evidence Processing} \label{EvidenceProcessing}

\begin{figure}[bh]
\includegraphics[width=\columnwidth]{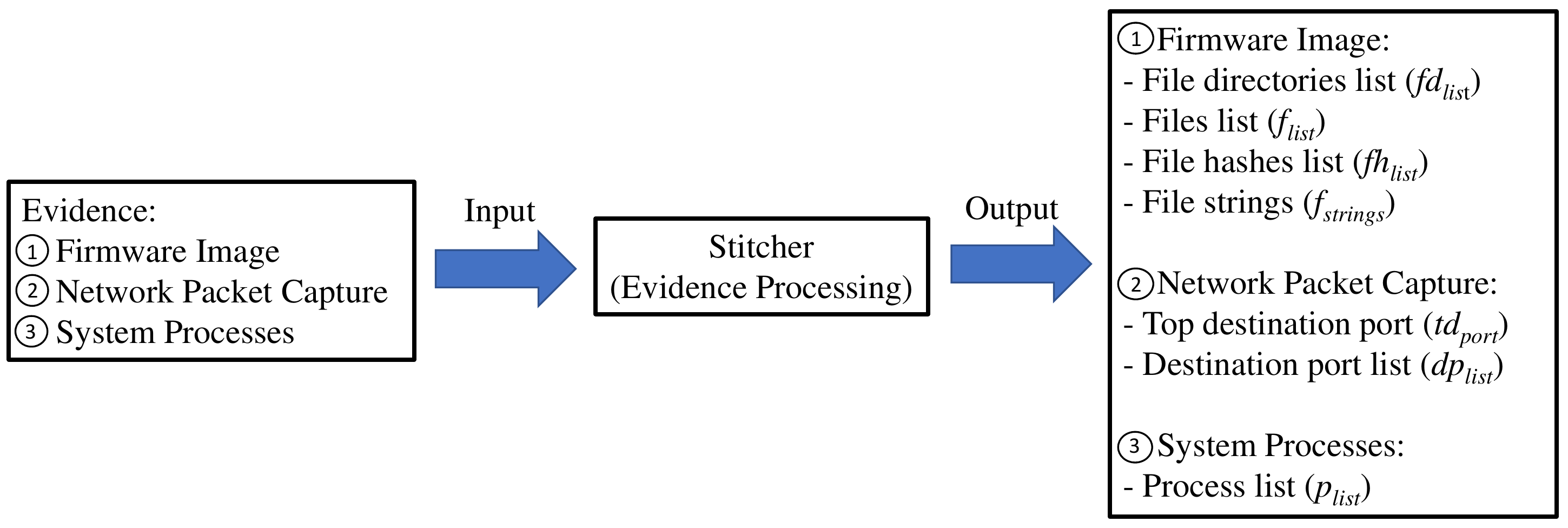}
\vspace{-18pt}\caption{Overview of Evidence Processing by \ST}
\label{fig:STEvidenceProcessing}
\end{figure}

In this section, we discuss how evidence is processed to facilitate correlation activities by \ST. Evidence processing is defined as preparing the evidence in a way that will allow an investigator to analyze the evidence. With reference to~\autoref{fig:STEvidenceProcessing}, 
the evidence that were processed come from three sources: firmware images, network packet capture and system processes. 

\paragraph*{\textbf{Firmware Image}}

\begin{figure}[th]
\includegraphics[width=\columnwidth]{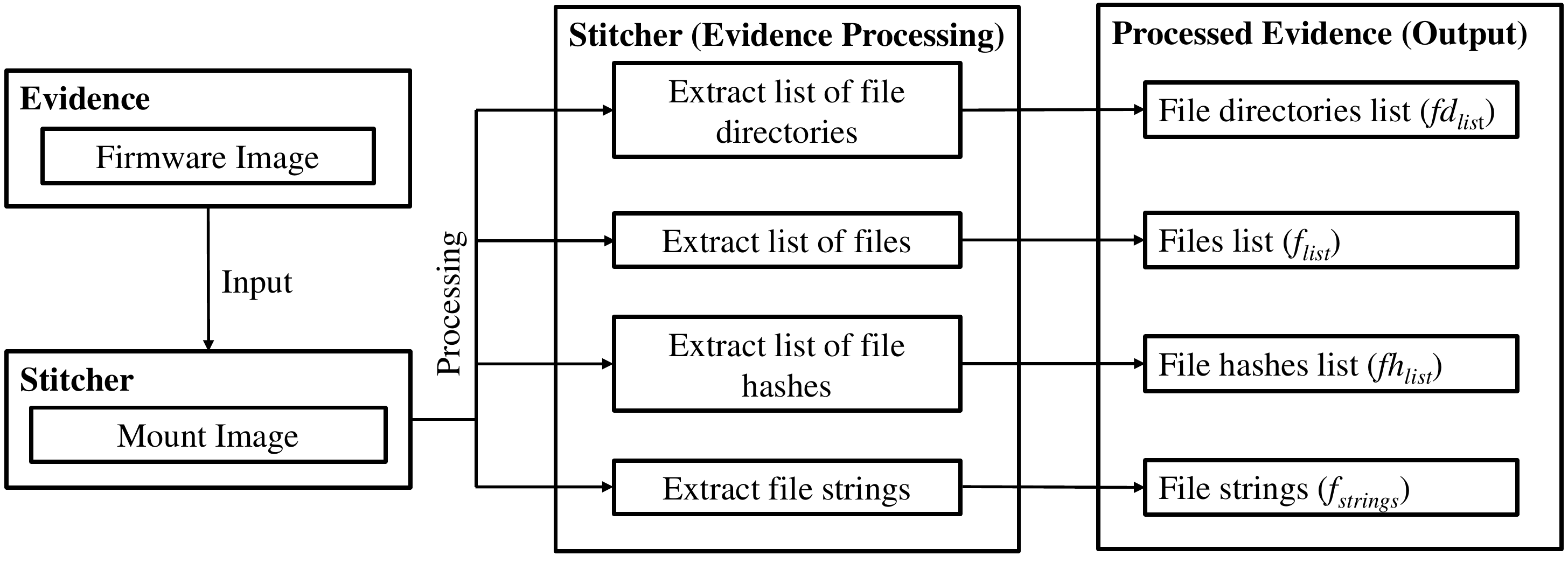}
\vspace{-18pt}\caption{Firmware Image Processing by \ST}
\label{fig:STFirmwareProcessing}
\end{figure}

\autoref{fig:STFirmwareProcessing} shows how the various processed evidence related to the firmware image are derived. We mount the disk image created by FAT after the firmware was virtualized. Following that, \ST processes the image to obtain the various artifacts such as file directories list (\textit{fd\textsubscript{list}}), list of files (\textit{f\textsubscript{list}}), list of file hashes (\textit{fh\textsubscript{list}}), and file strings (\textit{f\textsubscript{strings}}). The outputs are saved as text files to be used for correlation later. If there is a baseline or reference firmware image available, the same process is executed to obtain the baseline artifacts (denoted by \textit{bfd\textsubscript{list}}, \textit{bf\textsubscript{list}}, \textit{bfh\textsubscript{list}} and \textit{bf\textsubscript{strings}}).

\paragraph*{\textbf{Network Packet Capture}}

\begin{figure}[h]
\includegraphics[width=\columnwidth]{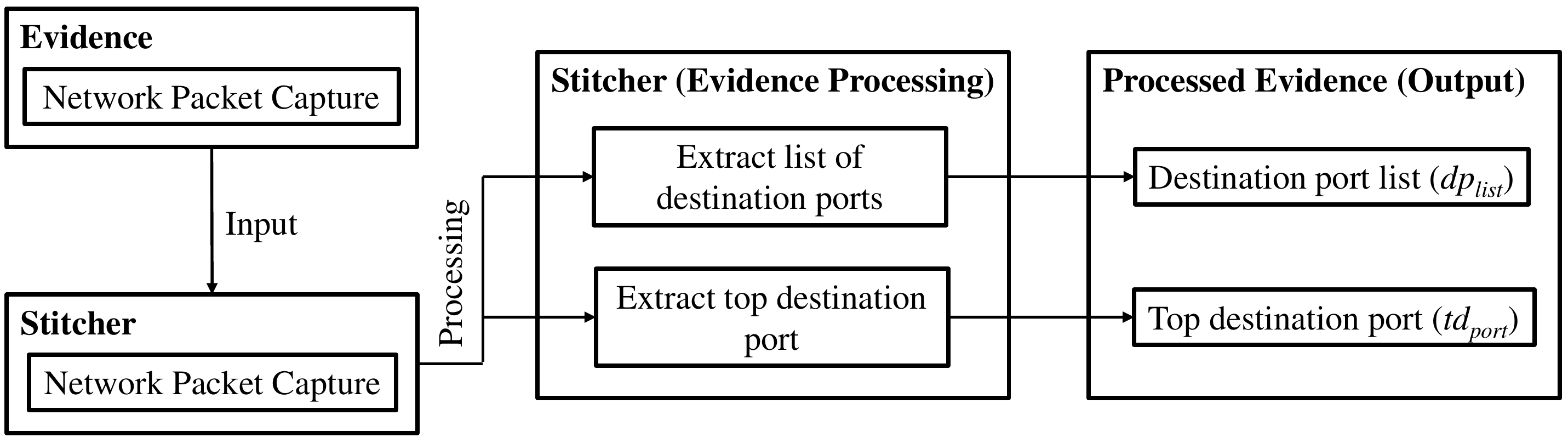}
\vspace{-18pt}\caption{Network Packet Capture Processing by \ST}
\label{fig:STNetworkProcessing}
\end{figure}

\ST also processes the network packet capture file to obtain the artifacts required for correlation later. With reference to~\autoref{fig:STNetworkProcessing}, the list of destination network ports (\textit{dp\textsubscript{list}}) and the top destination port (\textit{td\textsubscript{port}}) are generated from the source network packet capture. If there is a baseline or reference network packet capture available, the same process is executed to obtain the baseline network artifacts (denoted by \textit{bdp\textsubscript{list}} and \textit{btd\textsubscript{port}}).

\paragraph*{\textbf{System Processes}}
The list of system processes were originally in the form of a text file as they already have been retrieved. As such, they were the easiest to handle and no further processing was required. The system processes are denoted by \textit{p\textsubscript{list}} (with reference to~\autoref{fig:STEvidenceProcessing}) and baseline or reference system processes are denoted by \textit{bp\textsubscript{list}}.

\subsection{Evidence Correlation} \label{EvidenceCorrelation}

In this section, we explain how the evidence is correlated after being pre-processed. Evidence correlation is defined as identifying data points that appear consistently in evidence sources and such data points are related to each other. There are two possible types of evidence correlation, one is with just the gathered evidence. The other is with both the gathered evidence and baseline data provided by the manufacturer.

\subsubsection{Gathered evidence}

DFI have to examine and assess whether there is any evidence correlated with each other, and whether they consistently appear in each of the evidence source. Based on the attack scenario, DFI are expected to find the following pertinent evidence that will help them explain the scenario:

\begin{enumerate}[label=\roman*)]
    \item \textbf{Firmware.} DFI have to find that a new C program which was functioning as the backdoor, \textit{iSmartAlarmShell}, was inserted into the \textit{/sbin/} directory. When \textit{iSmartAlarmShell} is examined, it would yield text strings of a shell that is running over port 8888. Additionally, DFI have to identify that the string "iSmartAlarmShell" was also inserted in the \textit{rcS} file located inside the \textit{/etc\_ro/} directory.
	\item \textbf{Network.} DFI have to analyze the network packet capture file and identify that port 8888 was the most active network port due to the fact that the malicious attacker was using the port to send and receive data.
	\item \textbf{Process List.} DFI have to identify the presence of the process \textit{iSmartAlarmShell} that exists amongst the other legitimate processes that were running.
\end{enumerate}

Based on the above-mentioned artifacts, DFI have to manually investigate the relationships between processes and file names, processes and file strings and network port occurrences with file strings. However, \ST automates the investigation by correlating the artifacts to save time. With reference to Algorithm~\ref{Algo1}, Lines~\ref{PortFileStringStart} to~\ref{PortFileStringEnd} searches within the artifacts to correlate port numbers with file strings. Meanwhile, Lines~\ref{ProcessCompareStart} to~\ref{ProcessCompareEnd} searches within the artifacts to correlate process names with list of files and the respective file strings. 

\begin{algorithm} [t]
\caption{Evidence Correlation (Without Baseline)}
\label{Algo1}
\begin{algorithmic}[1]
\Require \textit{f\textsubscript{list}}, \textit{f\textsubscript{strings}}, \textit{dp\textsubscript{list}}, \textit{p\textsubscript{list}} // Processed evidence
\Ensure Correlation of evidence sources
\For{(each \textit{port}\ $\in \textit{dp\textsubscript{list}})$} // Iterate through list of ports and port number is stored as variable \textit{port} \label{PortFileStringStart}
    \If{($port\ \in\ f\textsubscript{strings}$)} // If port number matches file strings
        \State print \textit{port}
    \EndIf
\EndFor \label{PortFileStringEnd}
\For{(each \textit{process}\ $\in \textit{p\textsubscript{list}})$} // Iterate through list of processes and process is stored as variable \textit{process} \label{ProcessCompareStart}
    \If{($process \in f\textsubscript{list}\cap f\textsubscript{strings}$)} // If process name matches file name in file list and file strings
        \State print \textit{process}
    \EndIf
\EndFor \label{ProcessCompareEnd}
\State exit
\end{algorithmic}
\end{algorithm}

\subsubsection{Gathered evidence and baseline data}

\begin{algorithm}[b]
\caption{Evidence Correlation (With Baseline)}
\label{Algo2}
\begin{algorithmic}[1]
\Require \textit{f\textsubscript{list}}, \textit{fh\textsubscript{list}}, \textit{dp\textsubscript{list}}, \textit{p\textsubscript{list}}, // Processed evidence
\newline \textit{bf\textsubscript{list}}, \textit{bfh\textsubscript{list}}, \textit{bdp\textsubscript{list}}, \textit{bp\textsubscript{list}} // Processed baseline evidence
\Ensure Correlation of evidence sources with baseline sources
\If{($p\textsubscript{diff} \leftarrow p\textsubscript{list} \cap  bp\textsubscript{list} \neq \phi$)} // Check difference between process list and baseline \label{ProcessDiffStart}
    \State Print \textit{p\textsubscript{diff}}
\EndIf \label{ProcessDiffEnd}
\If{($f\textsubscript{diff} \leftarrow f\textsubscript{list} \cap bf\textsubscript{list}\neq \phi$)} // Check difference between file list and baseline \label{FileListDiffStart}
    \State Print \textit{f\textsubscript{diff}}
\EndIf \label{FileListDiffEnd}
\If{($fh\textsubscript{diff} \leftarrow fh\textsubscript{list} \cap bfh\textsubscript{list}\neq \phi$)} // Check difference between file hash and baseline \label{FileHashDiffStart}
    \State Print \textit{fh\textsubscript{diff}}
\EndIf \label{FileHashDiffEnd}
\If{($dp\textsubscript{diff} \leftarrow dp\textsubscript{list} \cap bdp\textsubscript{list}\neq \phi$)} // Check difference between port list and baseline \label{PortsDiffStart}
    \State Print \textit{dp\textsubscript{diff}}
\EndIf \label{PortsDiffEnd}
\State exit
\end{algorithmic}
\end{algorithm}

\begin{figure}[h]
\centering
\includegraphics[width=\columnwidth]{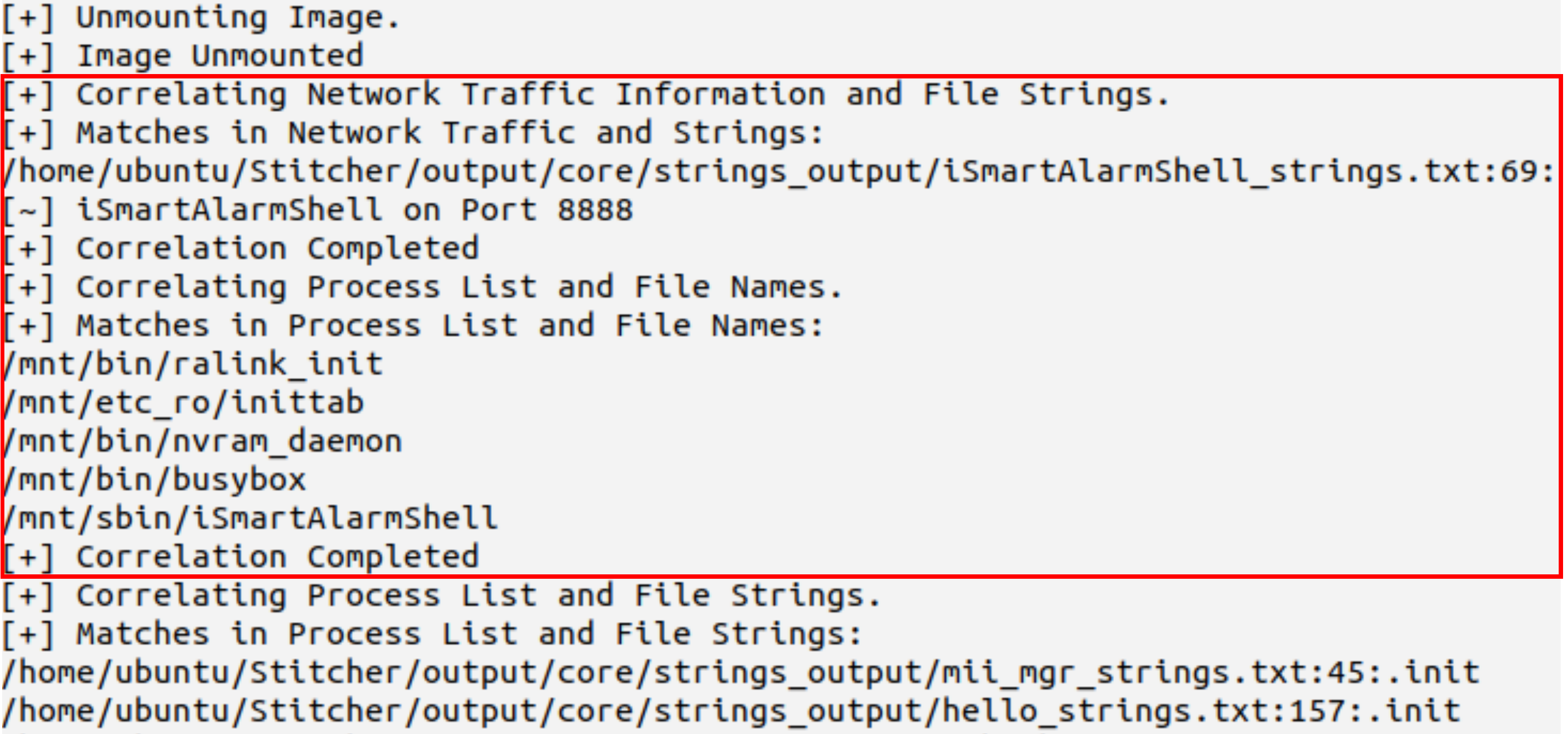}
\vspace{-18pt}\caption{Correlation of Evidence by \ST}
\label{fig:CorrelationST}
\end{figure}

The availability of a trusted reference point or baseline data, similar to a \textit{golden image}, can prove to be very helpful for DFI and even speed up the investigation process. Such data can usually be requested from the manufacturer when LEA or DFI request for assistance in solving cybercrime. Investigative workflows such as comparison of file hashes, file names, file strings and network communications between the IoT device being investigated on vis-a-vis a device provided by the manufacturer in an identical set-up can be done. If baseline data available, the baseline evidence will be processed in a similar fashion as stated in Section~\ref{EvidenceProcessing}. Algorithm~\ref{Algo1} will still be executed to correlate the scenario evidence, but following that, Algorithm~\ref{Algo2} will be executed to check for differences between processes (Lines~\ref{ProcessDiffStart} to~\ref{ProcessDiffEnd}), files (Lines~\ref{FileListDiffStart} to~\ref{FileListDiffEnd}), file hashes (Lines~\ref{FileHashDiffStart} to~\ref{FileHashDiffEnd}) and network ports (Lines~\ref{PortsDiffStart} to~\ref{PortsDiffEnd}) of scenario and baseline evidence. \autoref{fig:CorrelationST} further demonstrates the implementation of Algorithm~\ref{Algo2}.

\ST was made available in 3 different configurations for the purposes of the User Study. The configurations are as follows:
\begin{enumerate}
    \item \textbf{Configuration 1.} No tool is made available to a participant. Common open-sourced tools such as \textit{Wireshark}~\cite{Wireshark} and a specially crafted help file to offer some tips to participants were made available.
	\item \textbf{Configuration 2.} \ST with correlation functionality only.
	\item \textbf{Configuration 3.} \ST with full functionality (correlation and reporting).
\end{enumerate}
\section{Study Results} \label{StudyResults}
We present the findings of our user study in this section, and also provide insights to the data we obtained according to our Research Questions highlighted in Section~\ref{ResearchQn}.

\subsection{RQ1 Background Knowledge} \label{RQ1}

\begin{table}[h]
\caption{Background of DFI that Participated in User Study}
\begin{tabu} to 0.9 \columnwidth { | M{1cm} | m{6.8cm}| } 
 \hline
 \vspace*{4pt}\textbf{Category}\vspace*{4pt} & \vspace*{4pt}\hfil\textbf{Background}\vspace*{4pt} \\
 \hline
 A1 & \vspace*{4pt}Studied about digital forensics in an IHL \vspace*{4pt} \\
 \hline
 A2  & \vspace*{4pt}Took professional training and/or possess digital forensics certifications (e.g. GCFE, GCFA, GNFA, GREM, eCDFP, EnCE, ACE, etc)  \vspace*{4pt}\\
 \hline
 A3  & \vspace*{4pt}Investigated on cases where digital forensics was required \vspace*{4pt} \\
 \hline
 A4  & \vspace*{4pt}Studied about digital forensics in an IHL and investigated on cases where digital forensics was required  (A1 and A3)\vspace*{4pt}\\
 \hline
 A5  & \vspace*{4pt}Took professional training and/or possess digital forensics certifications (e.g. GCFE, GCFA, GNFA, GREM, eCDFP, EnCE, ACE, etc), and investigated on cases where digital forensics was required (A2 and A3)  \vspace*{4pt}\\
 \hline
 A6  & \vspace*{4pt}Studied about digital forensics in an IHL, took professional training and/or possess digital forensics certifications (e.g. GCFE, GCFA, GNFA, GREM, eCDFP, EnCE, ACE, etc), and investigated on cases where digital forensics was required (A1, A2 and A3) \vspace*{4pt}\\
\hline
\end{tabu}
\label{table:2}
\end{table}

We examine if a participant's background knowledge (respective categories of DFI user study participants outlined in~\autoref{table:2}) and years of experience made a difference when solving our scenario. Considering the usage of \ST may affect the outcome, we do not yet distinguish between the different tool configurations which could be an enabler in aiding DFI in solving the scenario. This is addressed in Section~\ref{RQ4} and Section~\ref{RQ5}.  

\begin{figure}[h]
\includegraphics[width=\columnwidth]{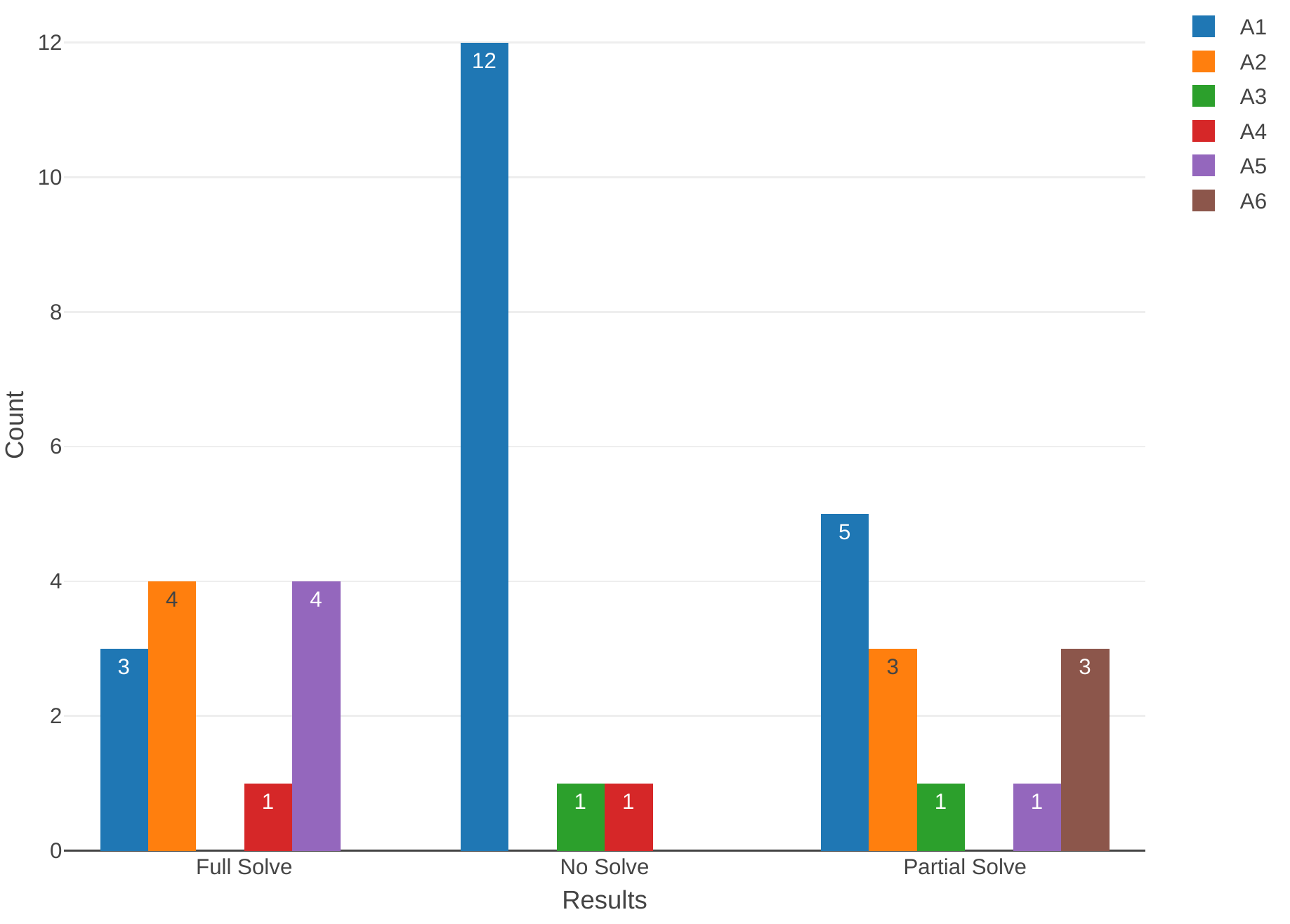}
\centering
\vspace{-18pt}\caption{Performance of DFI with respect to Background}\vspace{-8pt}
\label{fig:RQ1-DFI-Performance}
\end{figure}

\begin{figure}[h]
\includegraphics[width=\columnwidth]{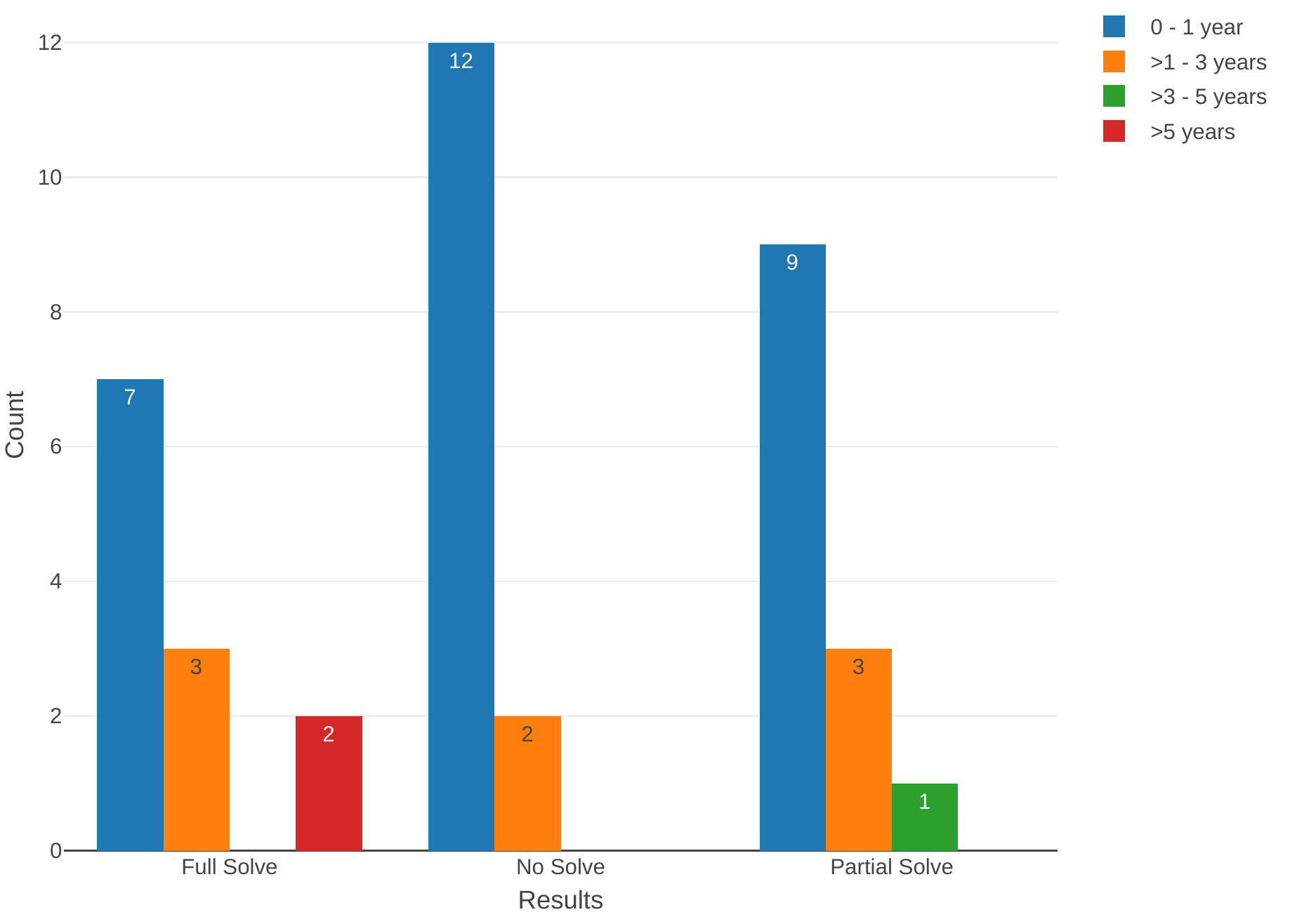}
\centering
\vspace{-18pt}\caption{Performance of DFI with respect to Years of Experience}\vspace{-6pt}
\label{fig:RQ1-DFI-Performance-Years}
\end{figure}

Based on the information provided by DFI during the survey, 
~\autoref{fig:RQ1-DFI-Performance} and~\autoref{fig:RQ1-DFI-Performance-Years} show 
how DFI performed in dealing with the scenario. In summary, 12 DFI achieved a full solve, 13 DFI achieved a partial solve, while 14 DFI could not solve the scenario within the given time limit. We observed that background knowledge does play a part in DFI performance. The majority of DFI (9 out of 12 DFI) who achieved a full solve had undergone professional training and/or worked on cases where digital forensics was required. Extremely experienced DFI ($>$5 years experience) were able to solve the scenario within the given time. 
Meanwhile,
most of the remaining DFI who had undergone professional training and/or worked on cases where digital forensics was required achieved a partial solve in the given scenario.

Another interesting observation is on 
DFI that belonged to Category A1 - DFI who only studied digital forensics in an IHL. More than half of DFI (12 out of 20) belonging to this category could not solve the scenario within the given time. This is due to the fact that IoT digital forensics is still an emerging area and IHL curriculum possibly do not cover IoT digital forensics. IHL digital forensics curriculum is also varied and it is possible that DFI learnt limited skills as they are ultimately constrained by the time and syllabus of the module. For example, certain IHLs may only teach theoretical aspects and assess knowledge via examinations, while other IHL assess student knowledge via hands-on examinations or projects. 

\subsection{RQ2 Challenges in Traditional Digital Forensics} \label{RQ2}

One of our contributions through this paper was to survey and present the current challenges faced by DFI when they had to carry out traditional digital forensics work. After some background research, we summarized the potential challenges in~\autoref{table:3}.

\begin{table} [h]
\centering
\caption{Challenges Faced by DFI in Traditional Digital Forensics}
\begin{tabu} to 0.9 \columnwidth{ | M{1cm} | m{6.8cm}| } 
 \hline
 \vspace*{4pt}\textbf{Category}\vspace*{4pt} & \vspace*{4pt}\hfil\textbf{Challenges in Traditional Digital Forensics} \vspace*{4pt}\\
 \hline
 B1 & \vspace*{4pt}Multiple evidence sources to examine \vspace*{4pt}\\
 \hline
 B2  & \vspace*{4pt}Multiple evidence sources to correlate activity \vspace*{4pt} \\
 \hline
 B3  & \vspace*{4pt}Creating an accurate and factual report  \vspace*{4pt}\\
 \hline
 B4  & \vspace*{4pt}Not enough training/knowledge\vspace*{4pt}\\
 \hline
 B5  & \vspace*{4pt}Multiple evidence sources to process \vspace*{4pt} \\
 \hline
 B6  & \vspace*{4pt}Classifying evidence to present them to client/report/court of law \vspace*{4pt}\\
 \hline
 B7 & \vspace*{4pt}Working with tight deadlines to solve cases in the shortest amount of time \vspace*{4pt} \\
 \hline
 B8 & \vspace*{4pt}Not enough manpower to work on cases \vspace*{4pt}  \\
 \hline
 B9 & \vspace*{4pt}Others \vspace*{4pt} \\
 \hline
\end{tabu}
\label{table:3}
\end{table}

\begin{figure} [t]
\includegraphics[width=\columnwidth]{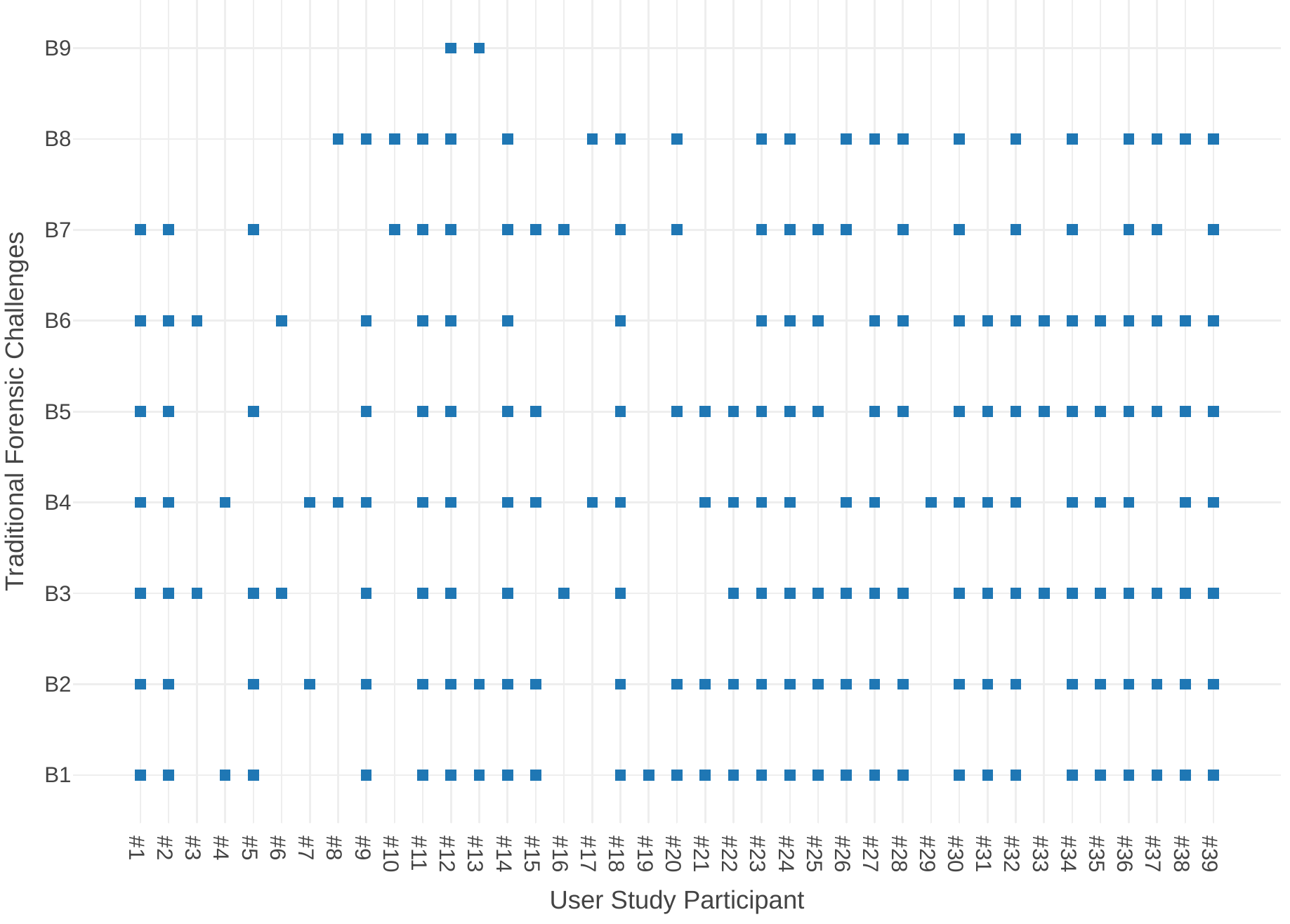}
\centering
\vspace{-18pt}\caption{Challenges Faced by DFI in Traditional Digital Forensics}\vspace{-5pt}
\label{fig:RQ2-DFI-Challenges-2}
\end{figure}

\begin{figure} [t]
\includegraphics[width=\columnwidth]{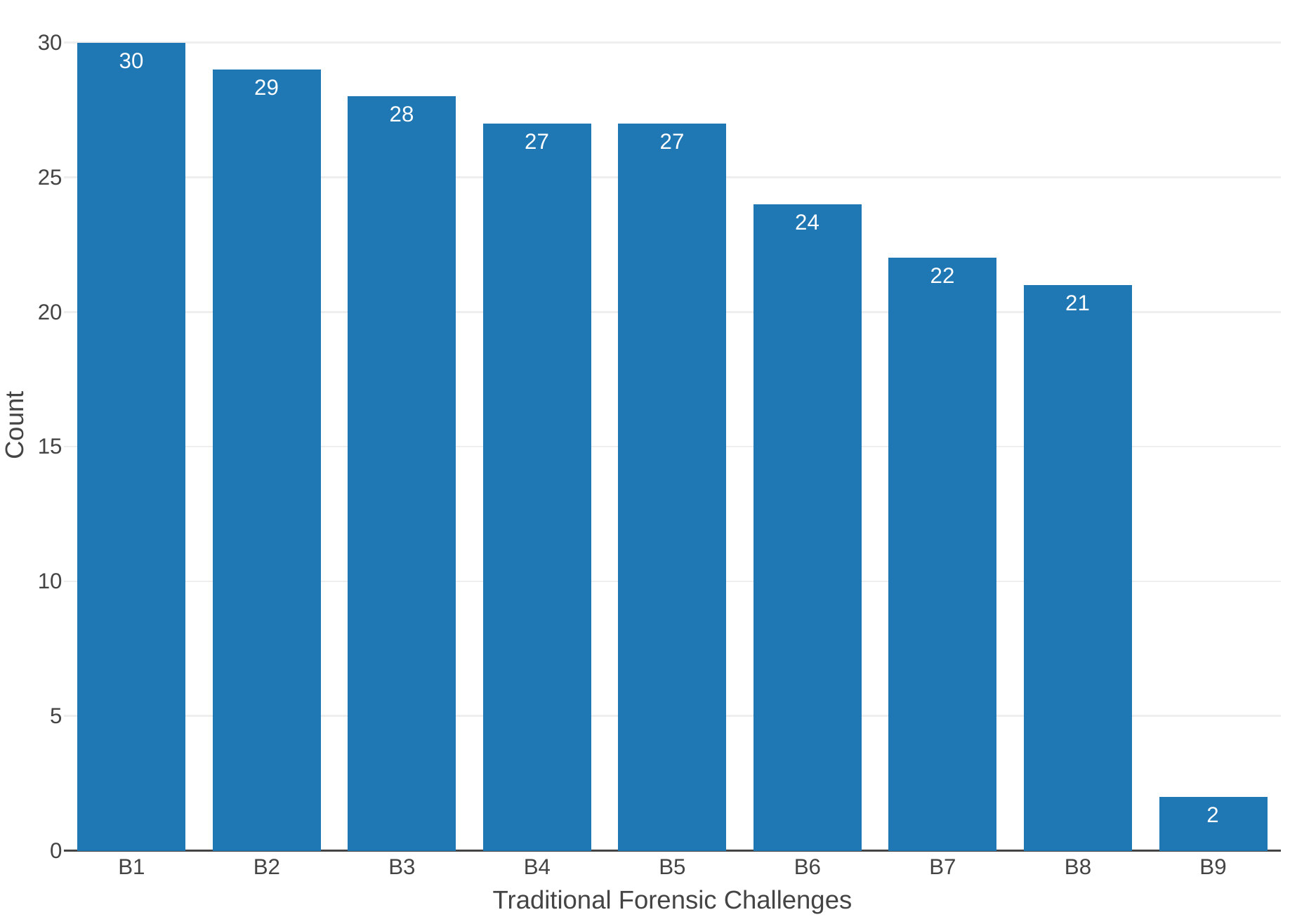}
\centering
\vspace{-18pt}\caption{Challenges Faced by DFI in Traditional Digital Forensics - Consolidated View}\vspace{-10pt}
\label{fig:RQ2-DFI-Challenges}
\end{figure}

We present the challenges faced by DFI individually based on their feedback in~\autoref{fig:RQ2-DFI-Challenges-2}, a consolidated view of the challenges is summarized in~\autoref{fig:RQ2-DFI-Challenges}. Most DFI had indicated that they faced more than 5 challenges in traditional forensics, except that DFI \#19 and \#29 selected only one challenge. For DFI \#12 and \#13 who selected B9 (Others), the challenges they faced were that there were no established general Standard Operating Procedures (SOP) for handling different types of cases they had to solve, such as malware and gathering of evidence. Another challenge they faced was understanding the environment where the case had happened.

The top 3 concerns highlighted by DFI during the study were multiple evidence sources to examine (B1), multiple evidence sources to correlate activity (B2) and creating an accurate and factual report (B3). This was in line with our goal of creating \ST, particularly for IoT digital forensics.

\subsection{RQ3 Challenges in IoT Digital Forensics} \label{RQ3}

\begin{figure} [h]
\centering
\includegraphics[width=0.9\columnwidth]{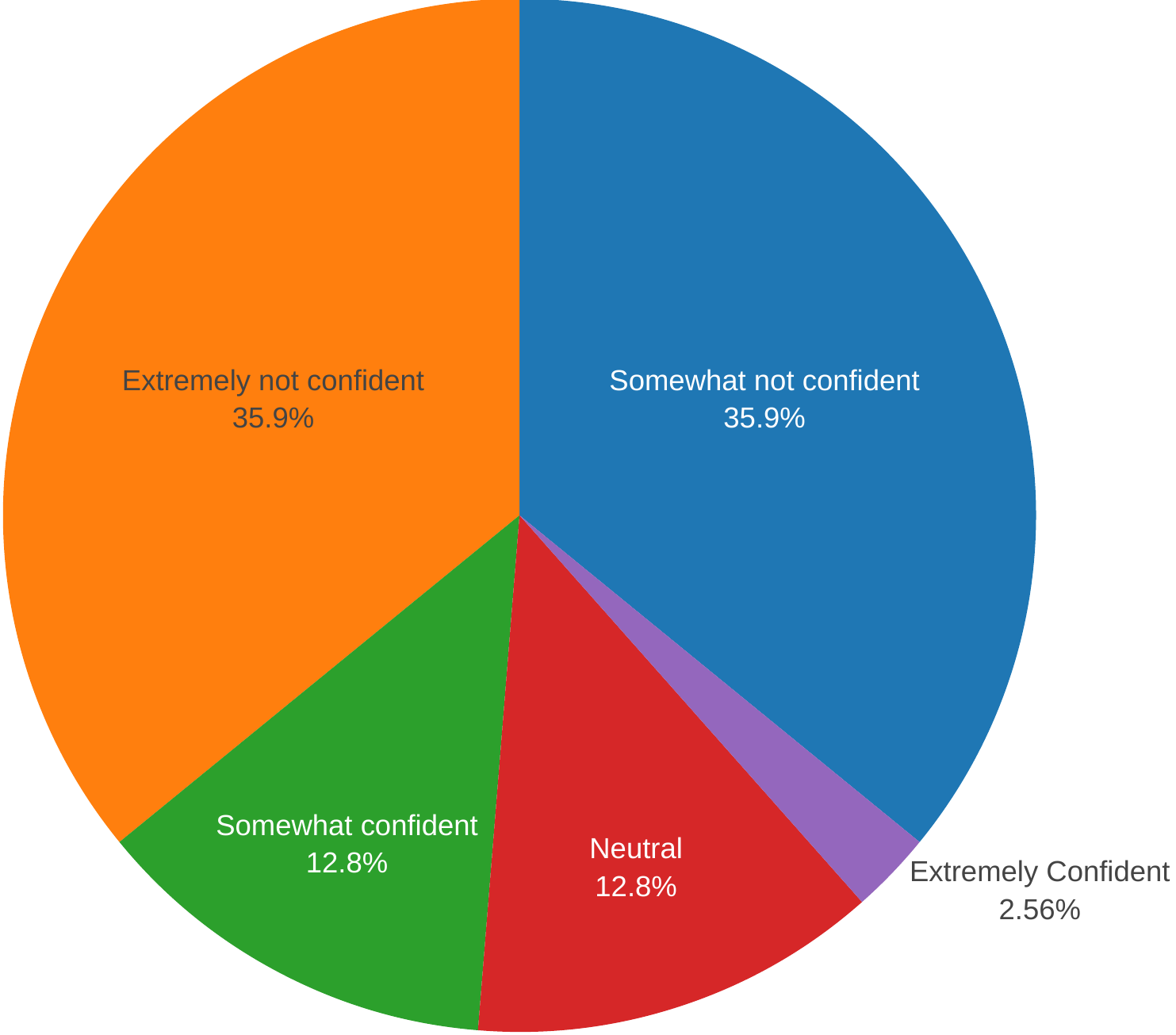}
\centering
\vspace{-7pt}\caption{DFI Confidence in Handling IoT Digital Forensics}\vspace{-10pt}
\label{fig:RQ3-DFI-Confidence}
\end{figure}

We also aim to provide another contribution -- an insight as to what challenges DFI are facing right now in the area of IoT digital forensics. As a litmus test, we asked the participants to rate their own confidence in handling IoT digital forensics (with reference to~\autoref{fig:RQ3-DFI-Confidence}) and only a mere 2.56\% of DFI indicated that they were extremely confident.

There are two sources of data that could be analyzed for this: challenges DFI thought they faced in IoT digital forensics in the broadest sense (cf. \autoref{table:4}) and the challenges DFI faced during the scenario (cf. \autoref{table:5}).

\begin{table} [h]
\centering
\caption{Challenges Faced by DFI in IoT Digital Forensics}
\begin{tabu}to 0.9 \columnwidth{ | M{1cm} | m{6.8cm}| } 
 \hline
 \vspace*{4pt}\textbf{Category}\vspace*{4pt} & \vspace*{4pt}\hfil\textbf{Challenges in IoT Digital Forensics}\vspace*{4pt} \\
 \hline
 C1 & \vspace*{4pt}Not enough training/knowledge\vspace*{4pt} \\
 \hline
 C2  & \vspace*{4pt}Potentially multiple evidence sources to examine\vspace*{4pt}  \\
 \hline
 C3  & \vspace*{4pt}Potentially multiple evidence sources to correlate activity\vspace*{4pt}  \\
 \hline
 C4  & \vspace*{4pt}Working with tight deadlines to solve cases in the shortest amount of time \vspace*{4pt}\\
 \hline
 C5  & \vspace*{4pt}Potentially multiple evidence sources to process \vspace*{4pt} \\
 \hline
 C6  & \vspace*{4pt}Creating an accurate and factual report \vspace*{4pt}\\
 \hline
 C7 & \vspace*{4pt} Classifying evidence (with respect to the context of IoT) to present them to client/report/court of law \vspace*{4pt} \\
 \hline
 C8 & \vspace*{4pt}Not enough manpower to work on cases  \vspace*{4pt}\\
 \hline
 C9 & \vspace*{4pt}Others \vspace*{4pt} \\
 \hline
\end{tabu}
\label{table:4}
\end{table}

\begin{table}[h!]
\centering
\caption{Challenges Faced by DFI During Scenario}
\begin{tabu} to 0.9 \columnwidth{ | M{1cm} | m{6.8cm}| } 
 \hline
 \vspace*{4pt}\textbf{Category}\vspace*{4pt} & \vspace*{4pt}\hfil\textbf{Challenges in Scenario} \vspace*{4pt}\\
 \hline
 D1 & \vspace*{4pt}Unsure of what to look out for\vspace*{4pt} \\
 \hline
 D2  & \vspace*{4pt}Unsure of the hypothesis (what had exactly happened) \vspace*{4pt} \\
 \hline
 D3  & \vspace*{4pt}Difficulty in correlating data points of interest \vspace*{4pt} \\
 \hline
 D4  & \vspace*{4pt}Unsure of what to include in report \vspace*{4pt}\\
 \hline
 D5  & \vspace*{4pt}Unsure of how to classify evidence sources for reporting and presentation to client/court of law  \vspace*{4pt}\\
 \hline
 D6  & \vspace*{4pt}Too many sources of evidence for analysis \vspace*{4pt} \\
 \hline
 D7 & \vspace*{4pt}There was not enough time to finish analysis of evidence  \vspace*{4pt}\\
 \hline
 D8 & \vspace*{4pt}Others  \vspace*{4pt}\\
 \hline
\end{tabu}
\label{table:5}\vspace{-6pt}
\end{table}

\begin{figure} [h]
\includegraphics[width=\columnwidth]{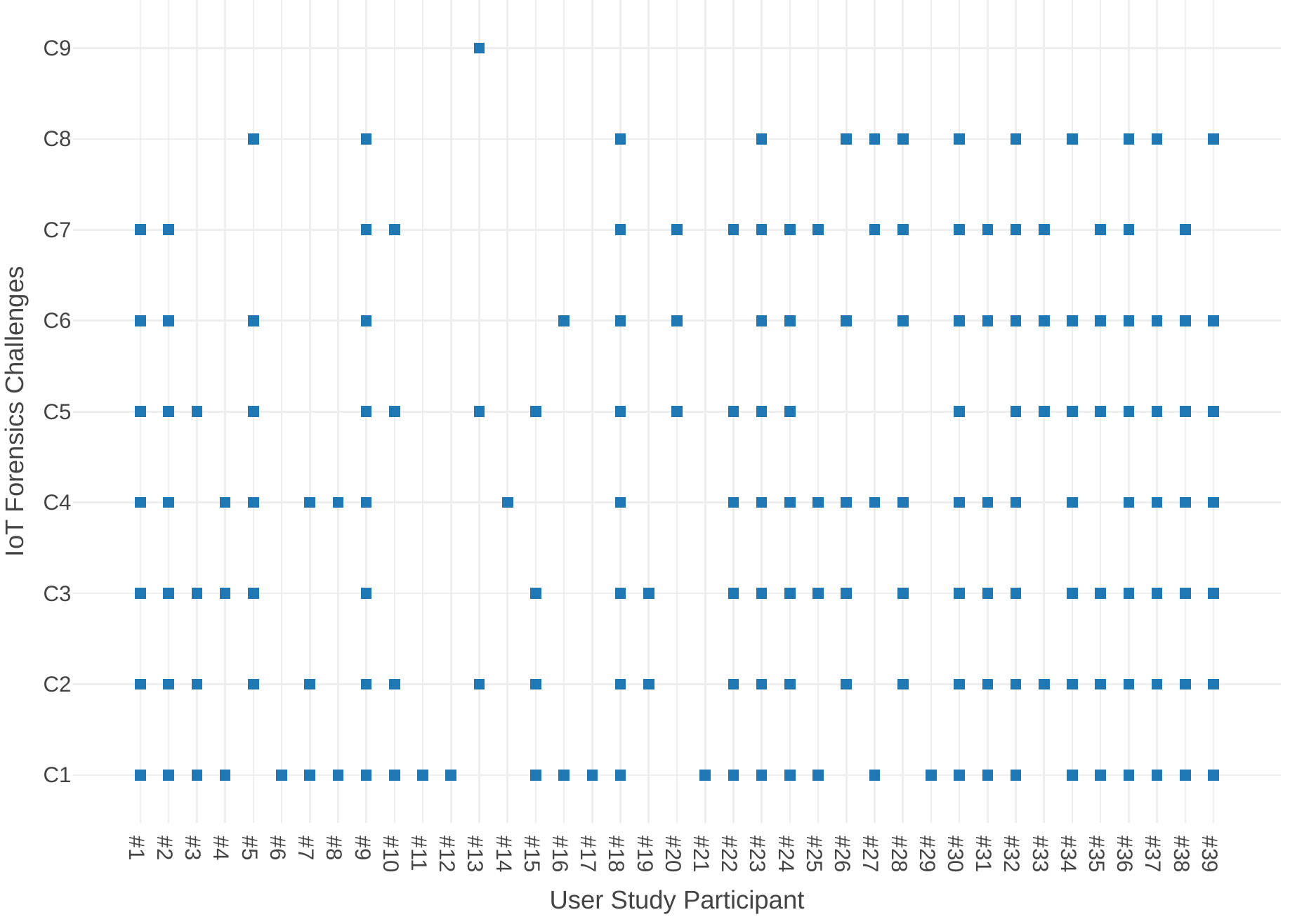}
\centering
\vspace{-18pt}\caption{Challenges Faced by DFI in IoT Digital Forensics}
\label{fig:RQ3-DFI-IoT-Challenges-2}
\end{figure}

\begin{figure} [h]
\includegraphics[width=\columnwidth]{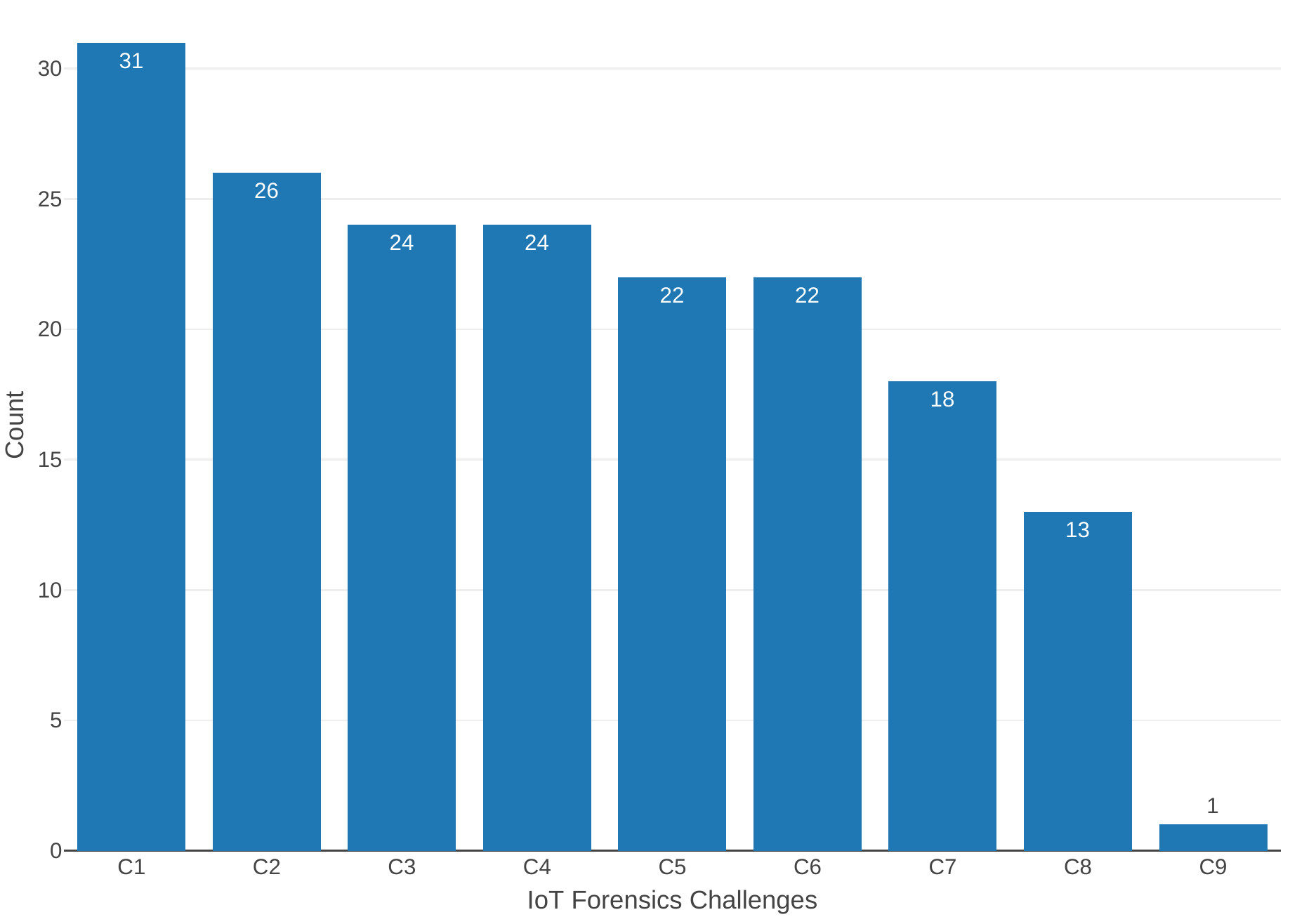}
\centering
\vspace{-18pt}\caption{Challenges Faced by DFI in IoT Digital Forensics - Consolidated View}\vspace{-10pt}
\label{fig:RQ3-DFI-IoT-Challenges}
\end{figure}

We present the challenges faced by DFI individually based on their feedback in~\autoref{fig:RQ3-DFI-IoT-Challenges-2}, and a consolidated view of the challenges is summarized in~\autoref{fig:RQ3-DFI-IoT-Challenges}. Only 6 out of 39 DFI (DFI \#6, \#11, \#12, \#17, \#21 and \#29) indicated only one challenge they would face in a potential IoT digital forensics case. Around half of the other DFI indicated that there were more than five challenges that they would face in a potential IoT digital forensics case. DFI \#13 selected ``Others'' (C9), and the challenge faced was being unable to understand the specifics of an IoT environment (such as logs not persisting after reboots and where data could be stored) which is similar to C1.

The top four challenges faced by DFI in IoT digital forensics were not enough training/knowledge (C1), potentially multiple evidence sources to examine (C2), potential multiple evidence sources to correlate activity (C3) and working with tight deadlines to solve cases in the shortest amount of time (C4) (both C3 and C4 had a tie of 24). It is not surprising to see C1 be the top challenge selected by DFI (31 out of 39). This also justifies why our research work is imperative and timely. As for C2, C3 and C4, they were more of operational challenges. A tool like \ST shall aid them in addressing such challenges.

\begin{figure} [h]
\includegraphics[width=\columnwidth]{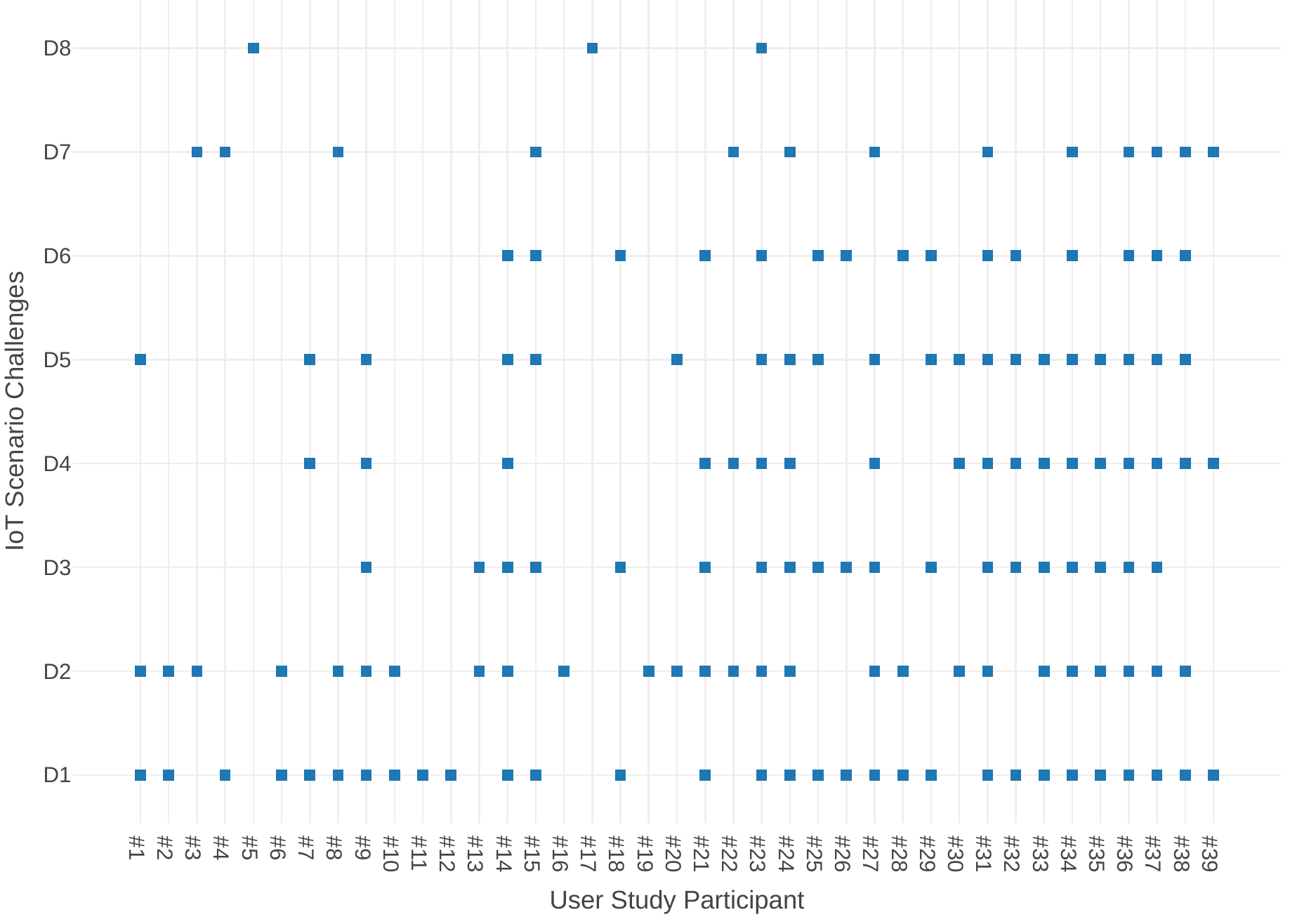}
\centering
\vspace{-18pt}\caption{Challenges Faced by DFI During Scenario}
\label{fig:RQ3-DFI-IoT-Scenario-Challenges-2}
\end{figure}

\begin{figure} [h]
\includegraphics[width=\columnwidth]{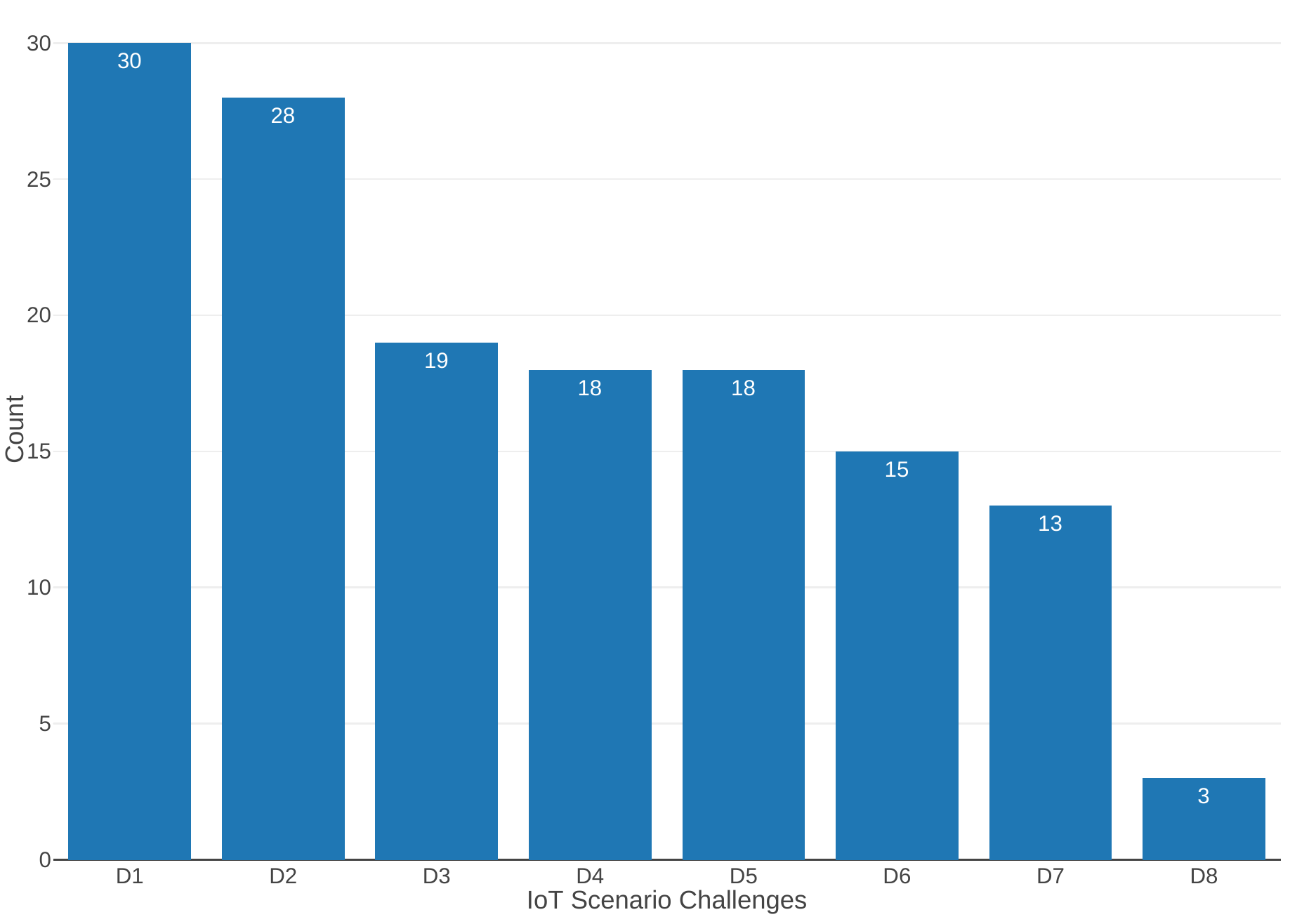}
\centering
\vspace{-18pt}\caption{Challenges Faced by DFI During Scenario - Consolidated View}\vspace{-10pt}
\label{fig:RQ3-DFI-IoT-Scenario-Challenges}
\end{figure}

The aforementioned challenges are ones that DFI
 {\em would} face in IoT digital forensics. We now examine what challenges DFI actually faced in the scenario and if they were correlated with each other. We present the actual challenges faced for each DFI in~\autoref{fig:RQ3-DFI-IoT-Scenario-Challenges-2}, and also provide a consolidated view of the challenges in~\autoref{fig:RQ3-DFI-IoT-Scenario-Challenges}.  

Fewer than half of the DFI faced more than five challenges. Only 6 out of 39 DFI (DFI \#5, \#11, \#12, \#16, \#17 and \#19) indicated only one challenge that they faced. DFI \#5, \#17 and \#23 selected ``Others'' (D8), and their challenges were being unable to interpret the evidence initially, no prior knowledge or training, and the inability to discern key data that were unique to IoT systems that could be used for digital forensics. DFI indicated fewer challenges per person in this question, and 13 out of 39 DFI felt that there was not enough time to finish the analysis of evidence. 

The top three challenges DFI faced during the scenario were being unsure of what to look out for (D1), unsure of the hypothesis (what had exactly happened) and difficulty in correlating data points of interest (D3). At a first glance, the top four challenges listed previously (C1, C2, C3 and C4) are different from D1, D2 and D3. However, after a closer look, we determined that they were interconnected. For example, a lack of training and knowledge certainly leads to 
an uncertainty
of what to look out for (C1 and D1) in IoT digital forensics cases. When there were multiple evidence sources to examine and correlate with tight deadlines, 
DFI found it hard to correlate the evidence sources and offer a correct hypothesis (C2, C3, C4, D2 and D3).

\subsection{RQ4 DFI Performance in IoT Digital Forensics} \label{RQ4}

\begin{figure} [h]
\includegraphics[scale=0.45]{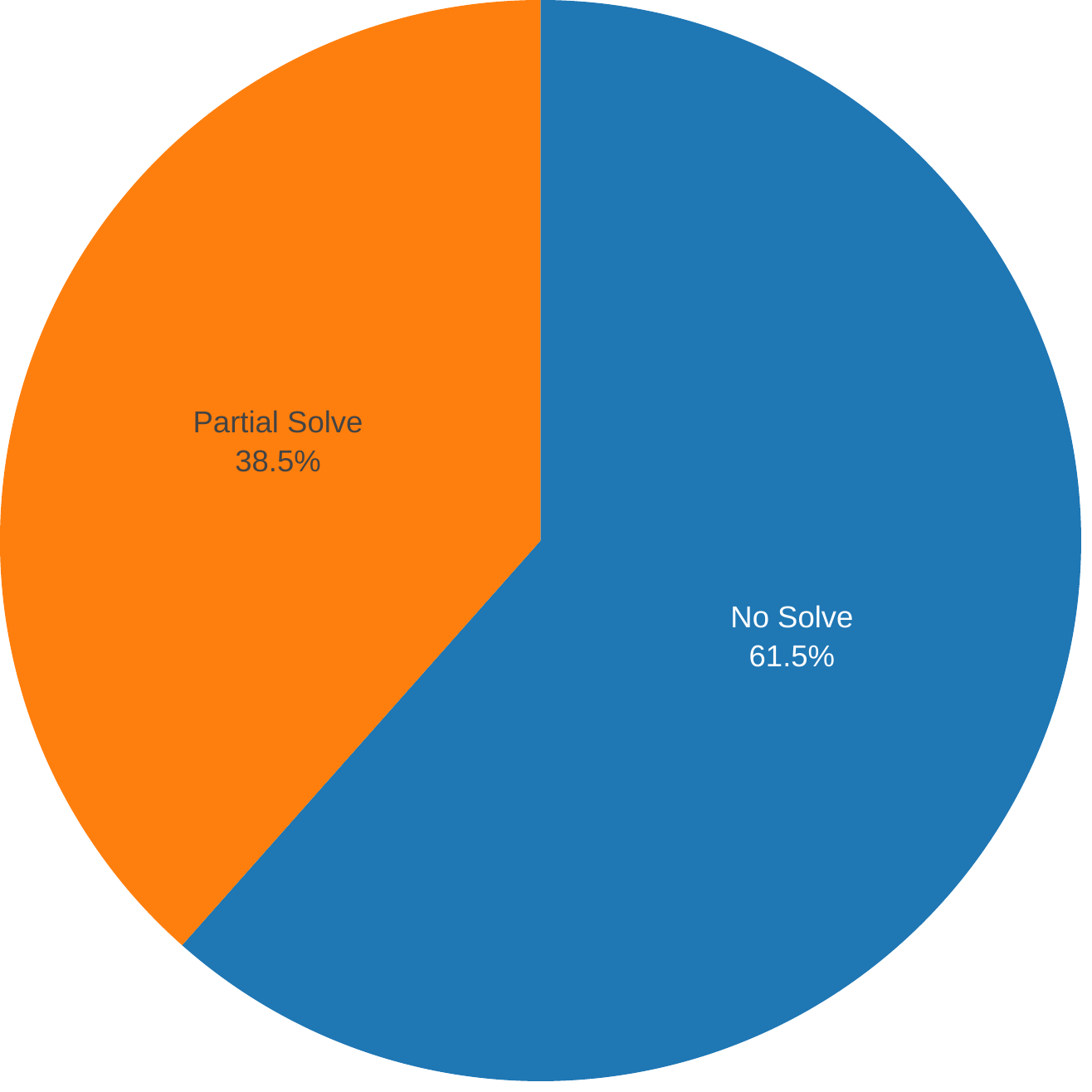}
\centering
\caption{Performance of DFI (Control Group)}\vspace{-10pt}
\label{fig:RQ4-DFI-Performance-Control-Group}
\end{figure}

In the experiment, we deliberately configured three versions of \ST as we wanted to investigate how DFI performed with currently available tools (the control group). With reference to~\autoref{fig:RQ4-DFI-Performance-Control-Group}, we observed that only 38.5\% (4 out of 13) of the DFI achieved a partial solve. The rest of DFI could not solve the scenario within the given time. This observation also corresponds to the challenges highlighted in Section~\ref{RQ3}.

\subsection{RQ5 Empowering DFI} \label{RQ5}

\begin{figure} [b]
\includegraphics[width=\columnwidth]{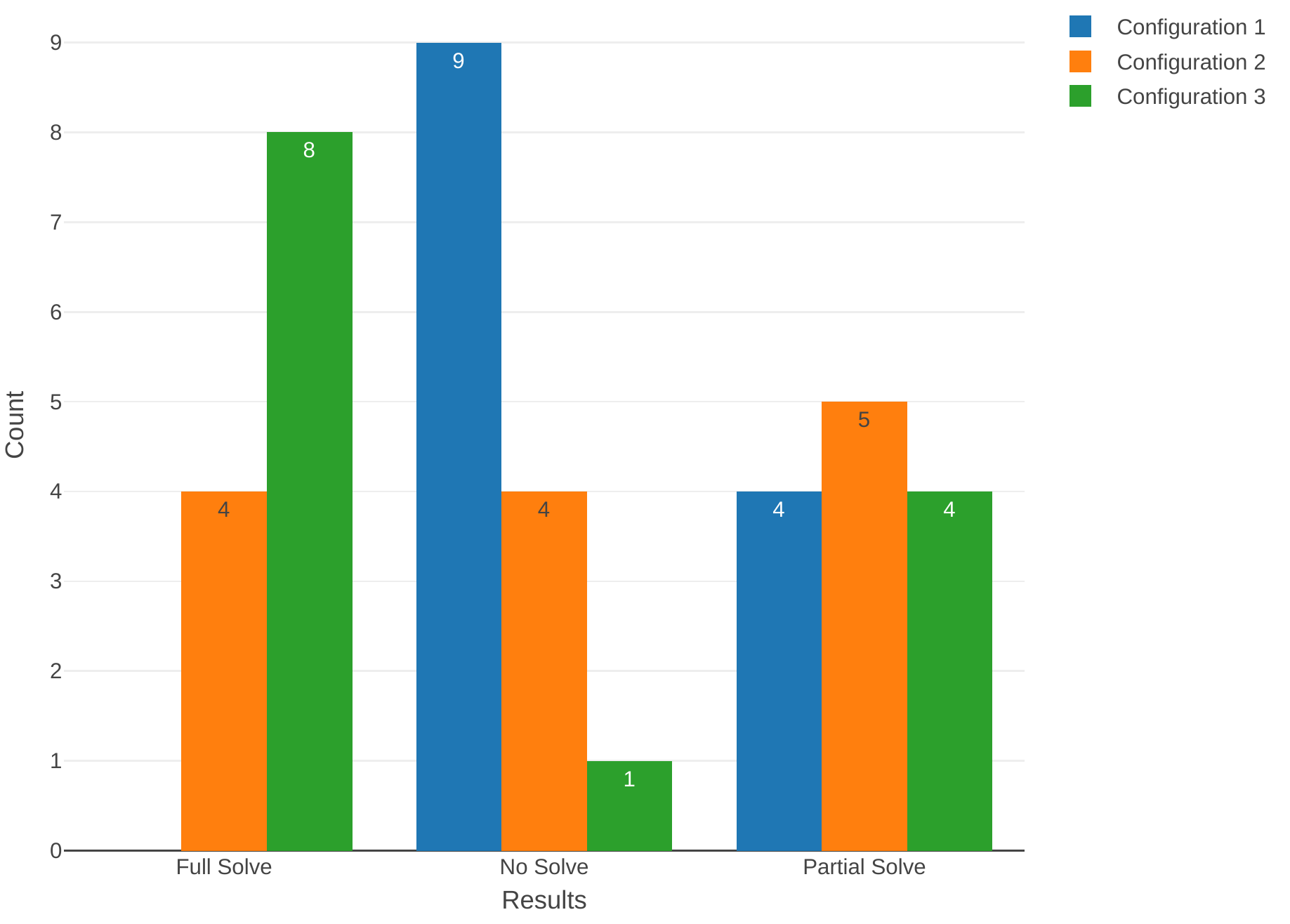}
\centering
\vspace{-18pt}\caption{Performance of DFI Using~\ST}
\label{fig:RQ5-DFI-Empowerment}
\end{figure}

\begin{figure}[h]
\includegraphics[width=\columnwidth]{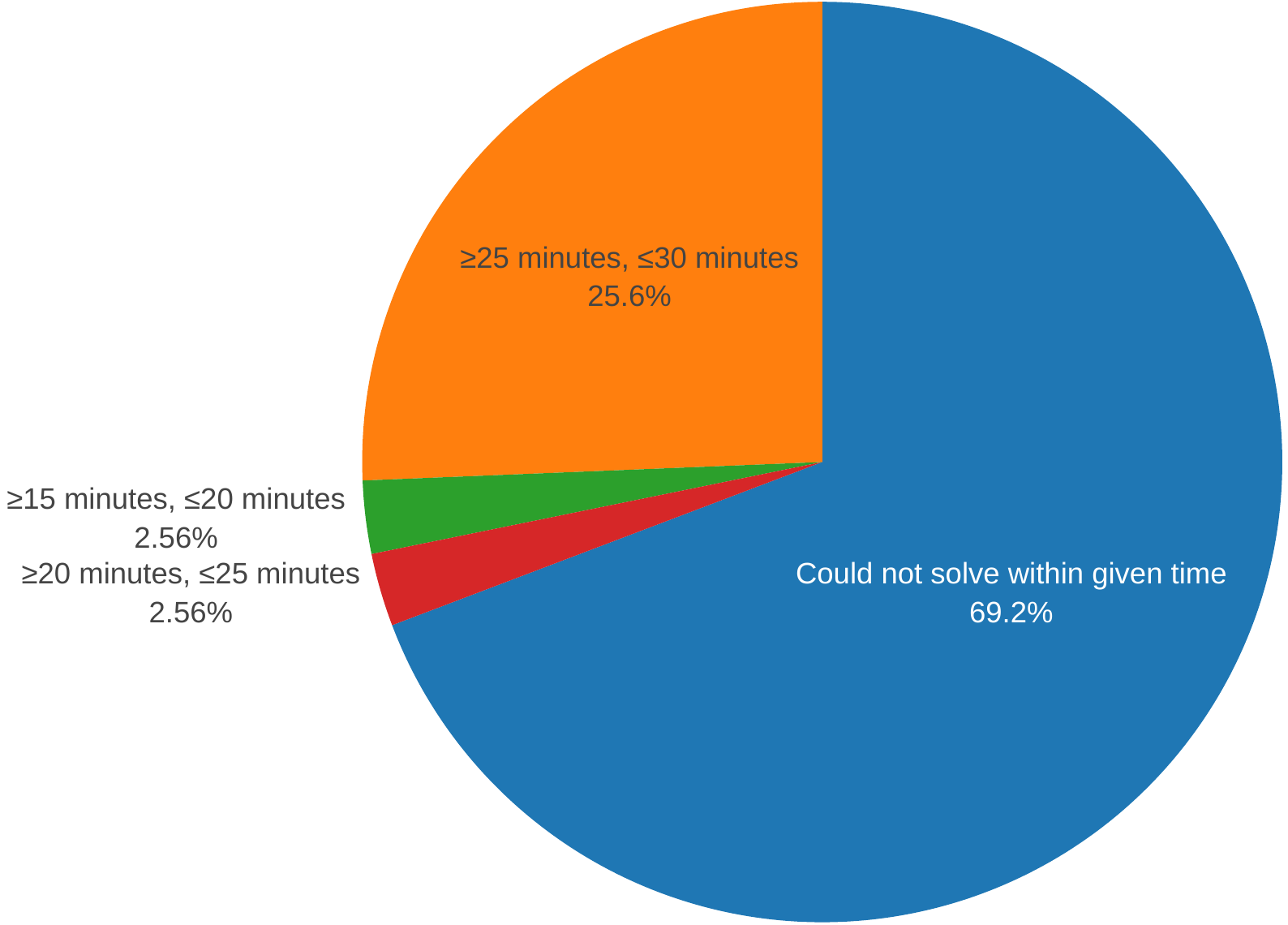}
\centering
\vspace{-18pt}\caption{Time Taken by DFI to Solve Scenario}
\label{fig:RQ5-DFI-Performance-Time-Taken}
\end{figure}

With regards to
the results presented in Section~\ref{RQ3} and Section~\ref{RQ4}, we now examine if the introduction of a software tool (\ST) designed to address the gaps could make DFI more efficient and effective in handling IoT forensics. From the feedback garnered from DFI, 96.2\% of users (25 out of 26) indicated that \ST truly assisted them in handling the scenario. With reference to~\autoref{fig:RQ5-DFI-Empowerment}, 61.5\% of users (8 out of 13) who used Configuration 3 achieved a full solve and 30.8\% of users (4 out of 13) who used Configuration 2 achieved a full solve. 
DFI who solved the scenario finished within 25 to 30 minutes (cf. \autoref{fig:RQ5-DFI-Performance-Time-Taken}), although 
some DFI
completed it in a shorter time. The fastest DFI completed the scenario in 15 to 20 minutes. This is because 
the DFI has prior knowledge or has researched on the area of IoT forensics.

Out of 26 DFI (excluding the 13 from the control group who used Configuration 1), 9 DFI achieved a partial solve. Although the partial solve does not capture solving the case entirely, these DFI could have potentially ended not solving the case at all without \ST. 

\begin{figure}

\includegraphics[scale=0.45]{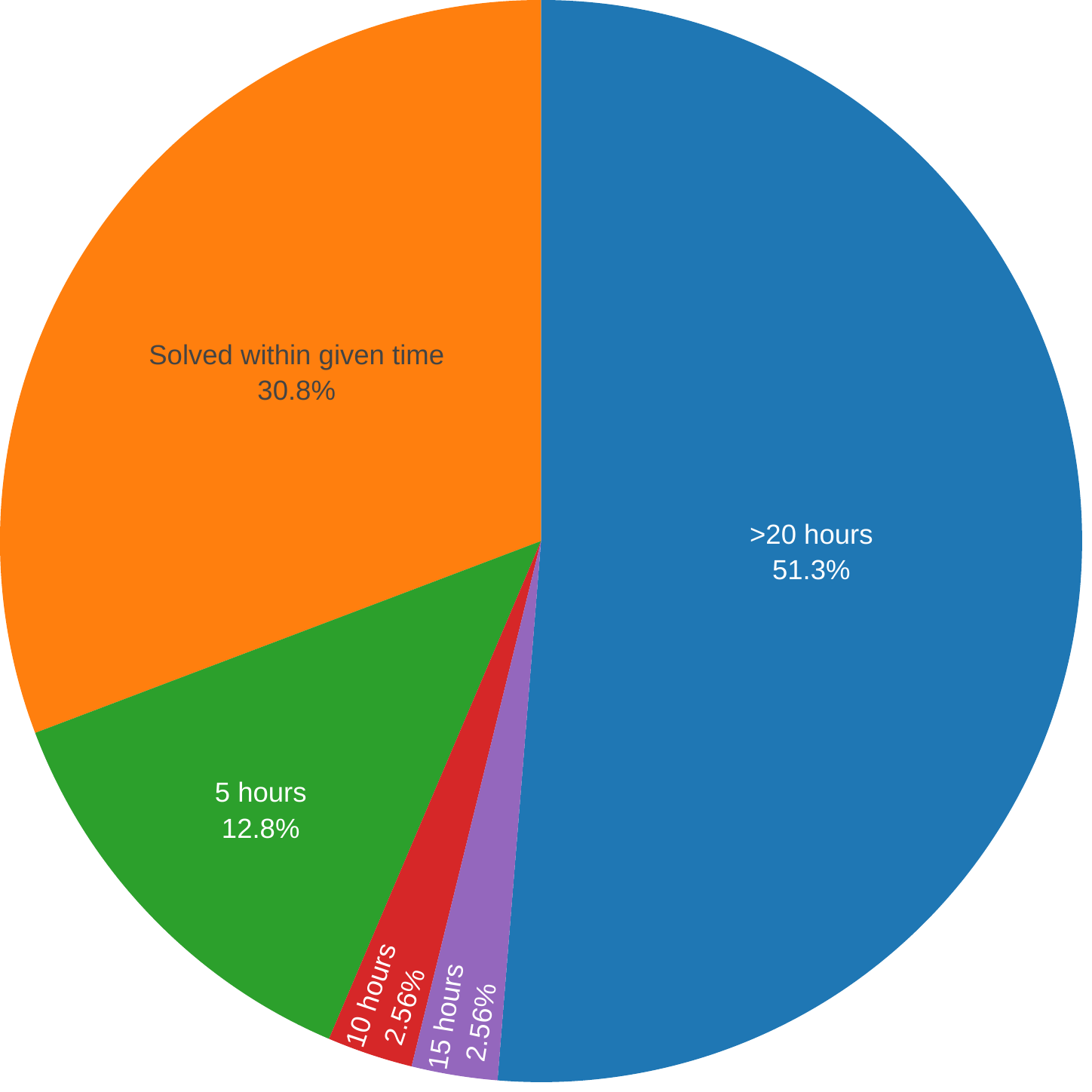}
\centering
\caption{Extra Time Needed by DFI to Solve Scenario}\vspace{-13pt}
\label{fig:RQ5-DFI-Performance-More-Time-Needed}
\end{figure}

We also surveyed the extra time potentially needed by DFI to solve the scenario (cf.~\autoref{fig:RQ5-DFI-Performance-More-Time-Needed}). More than half of the DFI indicated that they required more than 20 hours to solve the scenario. 
Note that solving of the scenario was limited to individual work and no team work was allowed, and 
that was one reason why DFI picked that choice. Usually, digital forensics would be a team effort as it is labour intensive and requires cross-checks. With more knowledge and training as well as a tool such as \ST, we envision that the time needed by DFI to handle IoT digital forensics cases would reduce. 
\section{Limitations} \label{Limitations}

Participants of the user study were only given 30 minutes to solve the scenario and an additional 10 minutes of writing time. This could have affected the results of participants solving the scenario, as given more time, those who partially solved the scenario could have solved it fully. However, the participants were working professionals and were unable to take part in a longer study. Additionally, some participants treated it as a new challenge to self-assess their skills and understanding, and thus did not mind the short amount of time given.

Despite the best efforts to gather a large group for the user study, we could only gather 39 local DFI from public and private sectors. Additionally, due to the constraints of monitoring user interaction with \ST, we did not consider seeking international DFI to participate in our user study. Thus, our data are skewed towards DFI based in Singapore and may not reflect the challenges faced by DFI in other countries. Although our data is region-specific, the study results and \ST could still be advisable to international DFI or LEA who are facing similar challenges in IoT digital forensics. 

\ST is able to work even in the absence of baseline evidence. However, it is possible that DFI will take a longer time for analysis of evidence if they do not have a baseline to refer to. \ST only correlates between the firmware image, network packet capture and system processes and highlight the data points that consistently appear, but DFI have to investigate if those were legitimate or malicious in nature. This issue can be addressed by requesting device manufacturers for the baseline evidence sources to assist in investigation. Alternatively, DFI or LEA can also consider working with device manufacturers to set up a facility or data repository of baseline evidence gathered from IoT devices. With the availability of a baseline reference point, DFI can identify malicious artifacts quickly. 

\ST only supports the classification and correlation of firmware images, network packet captures and system processes currently. Further work will be required to extend the classification and correlation of other types of digital evidence, such as digital evidence from the smartphone applications of IoT devices and data stored in cloud servers. Given more sources of evidence to correlate, DFI could obtain a more comprehensive output from \ST. However, evidence sources such as those from the cloud are usually harder and time consuming to obtain as they would require warrants or international cooperation from other LEA. Hence, \ST was designed with a focus on evidence sources that could be directly retrieved from the crime scene.

Finally, \ST is not yet compatible with other forensic tools, and also has not yet implemented the Cyber-investigation Analysis Standard Expression (CASE) and Unified Cyber Ontology (UCO) as proposed by~\cite{CASEY201714}. We want to address the capability gap experienced by DFI in terms of IoT forensics first, and leave the implementation of CASE and UCO in \ST as a future work.
\section{Related Work} \label{RelatedWork}

Some prior works on correlation in digital forensics have been explored. Chabot et al. presented a scenario reconstruction, semantic analysis and expert knowledge approach coupled with a formal-based timeline reconstruction and incident modelling~\cite{CHABOT2014S95}. Chabot et al. further suggested the reconstruction and analysis of incidents via an ontology-based approach~\cite{CHABOT201583}. These approaches work well for traditional investigations based on web browsing or even executable binaries on computers. However, their proposed correlation methods are not yet compatible for investigations based on IoT devices, particularly on digesting evidence sources such as firmware images and network packet captures. The methods also do not yet classify IoT evidence sources based on internationally agreed standards (such as ISO27050-1:2019 and ISO30141:2018) despite having their own proposed ontology. Finally, contrary to the aforementioned works, \ST has undergone trials with private and public sector DFI in a user study based on a realistic scenario emulating advanced adversaries. \ST was also well-received by the participants,  with 96.2\% of the users indicating that it aided them in solving the case.

Conventional methods such as memory forensics used by DFI on traditional computer systems are inapplicable to IoT devices. Case and Richard highlighted that ``it was difficult or impossible to acquire memory samples'' as such devices ran on a wide variety of custom operating systems and the hardware lacked memory forensics support~\cite{CASE201723}. The authors were also of opinion that in the event that memory forensics on such devices were possible, it would require the use of exploits in conjunction with tools and thus raising issues on the admissibility of such evidence in court~\cite{CASE201723}. In particular, we were also unable to extract any memory sample of iSmartAlarm CubeOne in our research, and agree that memory forensics on IoT devices cannot be executed currently.

A few solutions were proposed to address the challenges of IoT digital forensics. Amato et al. proposed the use of semantic technologies to correlate digital evidence and touts benefits such as integration of information, flexibility and classification~\cite{AmatoF2017Cdefi}. Kebande et al. proposed an integrated forensic investigation framework for IoT systems and incorporates several ISO standards in the framework~\cite{Kebande_2018}. Although the paper was comprehensive, it is not trivial to comply and implement such a framework in current environments which are already operational or already have had respective designs finalized. Unfortunately, neither solution offered any real-world implementation nor experiments were conducted to demonstrate its viability. As such, they remain as theoretical suggestions to address the challenges of IoT forensics.

New techniques such as using a multi-objective evolutionary algorithm to increase hardware security have been proposed to address and reduce the impact of sophisticated cyber-attacks~\cite{Marcelli:2017:EAH}. By securing hardware and reducing the attack surface of embedded systems, cybercriminals will face increased difficulty in exploiting IoT devices and thus IoT cybercrime should decrease, giving DFI and LEA a respite. Unfortunately, real-world implementations of such techniques by vendors have not been observed.

An interesting research conducted by Xu et al. explored the usage of attack trees to reconstruct the attack and gather relevant forensic evidence following the attack path~\cite{8823391}. This approach offers a potentially comprehensive coverage of forensic evidence and makes reporting intuitive. However, this method embraces several shortcomings. For example, if the proposed scenario is not applicable when compared against the root, the tree has to be rebuilt again to incorporate new scenarios and the corresponding evidence has to be acquired. This increases the workload for DFI with more time required for investigation. The generation of attack tree would require DFI to think like an attacker/criminal, which is at times challenging for DFI who are new to cybercrime related investigation work. In addition, the generation of attack trees in the context of IoT devices may inevitably point to the need for evidence from cloud service providers or servers located in geographically different areas with varying legal jurisdictions. Retrieval of such evidence are challenging for DFI/LEA, and decreases the effectiveness and efficiency of the proposed method.

In 2011, a user study was conducted by Hibishi et al. to investigate the usability of widely used commercial and open-source forensic software~\cite{5931114}. The authors found that users who indicated the various software in the survey often required a depth of technical knowledge before they could be used. Moreover, there were also conflicting preferences highlighted by the participants (a do-it-all tool vs. a tool that does a particular job extremely well). Our user study differs by focusing on a wider variety of challenges faced by DFI including and not limited to forensic tools. We provide an updated view of challenges faced by DFI in both traditional digital forensics and IoT digital forensics. We also show how \ST could help DFI in addressing the challenges faced in IoT digital forensics despite the majority not having much background in IoT digital forensics.

Servida and Casey investigated on possible forensics artifacts that could be retrieved from IoT devices and the corresponding smartphone applications used to control those IoT devices~\cite{SERVIDA2019S22}. The authors argued that further research was needed in home security systems, smart assistants and smart firewalls. They also demonstrated that there were forensic artifacts in IoT devices which could be admissible in court. Finally, the authors developed plugins to further extend \textit{Autopsy}, an open-source digital forensics tool. As current forensic tools for IoT do not yet classify and correlate IoT forensic evidence, we believe \ST fills this gap and would influence further research in the area of evidence correlation in IoT digital forensics.
\section{Discussion and Conclusion} \label{Conclusion}

We conducted a user study with 39 DFI from both public and private sectors to present the challenges DFI faced in traditional and IoT digital forensics. Based on the user study, we observed the following key findings:

\begin{enumerate}
    \item \textbf{Background Knowledge.} DFI who only studied digital forensics in an IHL (Category A1) did not perform well in the user study. More than half of DFI (12 out of 20) belonging to this category could not solve the scenario within the given time.
	\item \textbf{Challenges in Traditional Digital Forensics.} The top challenges highlighted by DFI were multiple evidence sources to examine, multiple evidence sources to correlate activity and creating an accurate and factual report.
	\item \textbf{Challenges in IoT Digital Forensics.} The top challenges faced by DFI in IoT digital forensics were not enough training/knowledge, potentially multiple evidence sources to examine, potential multiple evidence sources to correlate activity and working with tight deadlines to solve cases in the shortest amount of time.
	\item \textbf{DFI Performance in IoT Forensics.} DFI in the control group could not fully solve our simulated IoT crime. Only 38.5\% of DFI in the control group achieved a partial solve. For the rest of DFI using \ST, 96.2\% of DFI indicated that \ST assisted them in handling the crime and 61.5\% of DFI using \ST with its full features solved the crime completely.
\end{enumerate}

Our user study has demonstrated that there are multiple opportunities to enhance the state-of-the-practice in IoT digital forensics. The learning objectives, syllabus and assessment of digital forensics modules taught at undergraduate or postgraduate level could be reviewed to include guidance and concepts of IoT digital forensics so as to better prepare students for a new paradigm in digital forensics. The identified challenges in both traditional and IoT digital forensics can serve as potential research directions for researchers. Finally, we also attempted to address the consistency and correlation problem in digital forensics by creating a new tool named \ST to classify and correlate IoT forensic evidence. With the user study and the tool, we believe that we have lightened the burden of IoT investigation and challenges faced by DFI and LEA.

\section{Acknowledgements}

We would like to thank the pilot study candidate for assisting us in validating our user study setup, and the participants from both public and private sectors for volunteering their valuable time to participate in our user study and giving useful feedback on \ST. This work is partially supported by the Ministry of Education of Singapore under the grant MOE2018-T2-1-098.

{
	\bibliographystyle{unsrt}	
	\bibliography{references}
}


\end{document}